\documentclass[prd,reprint,onecolumn,eqsecnum,floats,aps,amsmath,amssymb,nofootinbib,superscriptaddress,showpacs]{revtex4-1}

\usepackage[utf8]{inputenc}
\usepackage[T1]{fontenc}
\usepackage[english]{babel}

\usepackage{amsmath,amsthm,amssymb,amsfonts}
\usepackage{tensor}
\usepackage{xcolor}
\usepackage{braket}

\numberwithin{equation}{section}
\numberwithin{thr}{section}
\numberwithin{chr}{section}
\numberwithin{df}{section}

\newcommand{\scpr}[2]{\langle#1\, \vert \, #2 \rangle}

\begin{document}

\title{Quantum theory of charged isolated horizons}

\author{Konstantin Eder}
\email{konstantin.eder@gravity.fau.de}
\affiliation{Department of Physics, Ludwig Maximilian University of Munich, Munich (Germany)}
\affiliation{Department of Physics, 
Friedrich-Alexander-Universit\"at  Erlangen-N\"urnberg (FAU), Erlangen (Germany)}
\author{Hanno Sahlmann}
\email{hanno.sahlmann@gravity.fau.de}
\affiliation{Department of Physics, 
Friedrich-Alexander-Universit\"at  Erlangen-N\"urnberg (FAU), Erlangen (Germany)}

\begin{abstract} 
We describe the quantum theory of isolated horizons with electromagnetic or non-Abelian gauge charges in a setting in which both gauge and gravitational field are quantized. We consider the distorted case, and its spherically symmetric limit. 
We show that
the gravitational horizon degrees of freedom give rise to the Bekenstein-Hawking relation, with lower order terms giving some corrections for small black holes. We also demonstrate that one can include matter degrees of freedom into the state counting. We show that one can expect (potentially divergent) contributions proportional to the area, as well as logarithmic correction proportional to the horizon charge. This is qualitatively similar to results on matter contributions obtained with other methods in the literature. 
\end{abstract} 
\maketitle


\tableofcontents

\section{Introduction} 
\label{se_intro}
The discovery of black hole thermodynamics \cite{Bardeen:1973gs}, and in particular the relationship of the entropy of a black hole with its area \cite{Bekenstein:1973ur} has led to efforts to pinpoint the microscopic degrees of freedom responsible in the existing theories of quantum gravity. 

In loop quantum gravity (LQG), an intriguing picture has emerged. One-dimensional excitations pierce the black hole horizon and endow it with area. The resulting punctures carry internal states which, when taken together with the different configurations of punctures account for the BH entropy. \cite{Smolin:1995vq,Rovelli:1996dv,Ashtekar:1997yu,Ashtekar:2000eq,Kaul:2000kf,Domagala:2004jt,Corichi:2006wn,Agullo:2008yv, Engle:2010kt,Ghosh:2011fc,Frodden:2012dq,Frodden:2012en,Ghosh:2013iwa}. The picture also generalizes to higher dimensions and to more general gravity theories 
\cite{Bodendorfer:2013jba,Bodendorfer:2013wga,Bodendorfer:2013sja}. 
For recent reviews see \cite{DiazPolo:2011np,Perez:2017cmj}

LQG is based on Ashtekar's discovery \cite{Ashtekar:1986yd}, later refined by Barbero \cite{Barbero:1994ap}, that the phase space for gravity can be embedded into that of SU(2) Yang-Mills theory. This is a step towards unification, as the kinematical description of gravity becomes that of a gauge field. Thus one would expect interesting effects when considering the combined system of gravity and gauge fields. Indeed this is the case as the results of Thiemann \cite{Thiemann:1997rt} on the finiteness of the quantum Hamiltonian constraint show. 

In these variables for gravity, the electric part of the field strength is responsible for spatial geometry. In particular, area is electric flux, 
\begin{equation}
a_{S}=\int_{S}{\underset{\raisebox{1pt}{$\Leftarrow$}}{{E}}}. 
\end{equation}
In the quantum theory, the one-dimensional excitations are flux tubes, carrying (Planck-) units of area.  

In the present work, we will bring together the quantum theory of matter and black holes, by considering the quantum theory of isolated horizons in the presence of YM gauge fields. 
This topic is not completely new: Classical aspects have been discussed for electromagnetic field already in \cite{Ashtekar:1999wa} and, more recently for YM theory in \cite{Corichi:2000dm}. 
However, for reasons not entirely clear to us, the quantum theory for this situation seems to not have been considered before. \cite{Ashtekar:2000eq} contains an argument that considering the quantum theory of the combined system is irrelevant in terms of determining black hole entropy for spherically symmetric horizons: The matter fields on the horizon are completely specified by the gravitational ones in this case, and can thus be eliminated classically.  Furthermore, there is no surface term of the matter symplectic structure, and hence no motivation to consider surface degrees of freedom of the matter fields separately in the quantum theory. This argument may be correct in terms of entropy alone, and for the spherically symmetric case only. However, in subsequent publications, it has sometimes been taken to mean that considering the combined quantum theory is wholly unnecessary, or that it has already been taken care of. This is not the case. 

It is true that \emph{on the horizon} the matter fields are completely specified by the gravitational ones through the boundary conditions in the spherically symmetric case. However, this obviously ceases to be true away from the horizon. There the fields live their separate lives. Furthermore, it is nor true for general isolated horizons. 

One knows quite well how quantum states of Yang-Mills fields look in LQG \cite{Corichi:1997us,Thiemann:1997rt}: They form the same type of one-dimensional excitations as the gravitational ones, and give corresponding contributions to the 
electric flux 
\begin{equation}
Q_{S}=-\frac{1}{4\pi}\int_{S}{\underset{\raisebox{1pt}{$\Leftarrow$}}{\mathbf{E}}}
\end{equation}
through a surface $S$. For a charged horizon $H$, its electric charge can be obtained as the flux $Q_{H}$. One must conclude that when the gauge field is quantized along with the gravitational field, 
the excitation of the gauge field must pierce the horizon and endow it with charge in exactly the same way as the gravitational excitations do to create its area $A_H$. The boundary conditions tying the matter fields to the gravitational one  then translate to conditions on the ``area-'' and Yang-Mills charges carried by the states into the punctures. 

Because of this coupling at the horizon, the first and most fundamental question that we intend to answer is: 
\begin{itemize}
\item Is the inclusion of quantized YM fields compatible with the BH picture, in the sense that there are enough states that satisfy the quantum boundary conditions to account for the BH entropy. 
\end{itemize}
As it turns out, this is by no means a trivial question. In particular, we will see that there are particular problems with the spherically symmetric case. 

If the answer is affirmative, a generalization of the picture of a quantized isolated horizon to a larger class of black holes has been obtained. This is a worthwhile goal in itself. But, once we have a consistent picture including quantized gauge fields, there are further interesting questions to be asked. For one thing, it has been variously suggested that matter fields do contribute to -- or are even the main ingredient of -- BH entropy, be it through their entanglement with the inside of the black hole, or otherwise. For one thing, it can be shown that entanglement entropy across boundaries in matter ground states is proportional to area (albeit divergent) 
\cite{Bombelli:1986rw,Srednicki:1993im}. On the other hand, it has been shown that matter leads to subleading corrections in the entropy at one loop order in path integral calculations, see for example \cite{Sen:2012dw} for the case 
of 4-dimensional Einstein gravity. Therefore we would like to ask:
\begin{itemize}
\item Is it possible to including the gauge field degrees of freedom on the horizon in the statistical ensemble, and what are the resulting corrections?
\end{itemize}
There are a number of other questions that can be addressed once a viable quantum theory of charged isolated horizons becomes available. One is the generalization and testing of the hypothesis that the quasinormal mode spectrum of Schwarzschild black holes is related to area spectrum in LQG. While we will briefly discuss some preliminary results on this here, we will report details on this elsewhere, see also \cite{KE_master}.  

The article is structured as follows: In the next section, we discuss the classical theory of type III isolated horizons in the presence of YM fields and charges, and define the corresponding phase space. In section \ref{se_typeIII} we describe its quantization, and the solutions of the boundary conditions. In section \ref{se_entropy}, we compute the horizon entropy under various assumptions. 
We summarize and discuss the results in section \ref{se_summ}. 

We work with metric signature $(-,+,+,+)$. The Yang-Mills structure group $G$ is assumed to be either $G=\mathrm{U}(1)$ or compact and semisimple throughout. Some details regarding spinor calculus and Newman-Penrose formalism can be found in appendix \ref{ap_spinor}.  

\section{Classical theory and phase space} 
\label{se_class}
\subsection{Canonical formulation of Yang-Mills theory}
We consider Yang-Mills theory as a non-Abelian gauge theory with semisimple compact structure group $G$ on a globally hyperbolic Lorentzian spacetime manifold $(M,g)$. Hence, the gauge potential $A$ of the Yang-Mills field is given by a certain $\mathfrak{g}$-valued connection 1-form on a $G$-principal bundle over $M$. Correspondingly, the field strength tensor is given by the associated curvature $F(A)^K=\mathrm{d}A^K+\tensor{f}{_{IJ}^K}A^I\wedge A^J$ where $\tensor{f}{_{IJ}^K}$ are structure coefficients of the Lie algebra. The action functional of the underlying field theory is then defined as follows 
\begin{equation}  
S_{\mathrm{YM}}[A]=\frac{1}{8\pi g^2}\int_{M}{\braket{F\wedge\ast F}}=\frac{1}{16\pi g^2}\int_M{\sqrt{-g}\tensor{F}{^{\mu\nu}^I}\tensor{F}{_{\mu\nu}^I}\,\mathrm{d}^4x}
\label{eq:II.1}
\end{equation} 
where $g$ is the Yang-Mills coupling constant. Here, we have chosen an arbitrary Ad-invariant bilinear form $\braket{\cdot,\cdot}$ on $\mathfrak{g}$ and extended it to $\Omega^{*}(M,\mathfrak{g})$ by setting $\braket{\alpha\wedge\beta}:=\alpha^I\wedge\beta^J\braket{\tau_I\otimes\tau_J}:=\alpha^I\wedge\beta^J\braket{\tau_I,\tau_J}$ for any $\alpha,\beta\in\Omega^{*}(M,\mathfrak{g})$ and a basis $\{\tau_I\}$ of $\mathfrak{g}$. In order to derive the corresponding Yang-Mills equations, we vary the Yang-Mills functional w.r.t. to the gauge field $A$. Using $\delta F(A)=\mathrm{d}_A\delta A$ for any variation $\delta A$ as well as $\braket{\mathrm{d}_A\omega\wedge\ast\eta}=\mathrm{d}\braket{\omega\wedge\ast\eta}+\braket{\omega\wedge\mathrm{d}_A(\ast\eta)}$ for any 1-form $\omega$ and 2-form $\eta$, we find
\begin{align}
\delta S_{\mathrm{YM}}&=\frac{1}{8\pi g^2}\int_{M}{\braket{\mathrm{d}_A\delta A\wedge\ast F}+\braket{F\wedge\ast\mathrm{d}_A\delta A}}=\frac{1}{4\pi g^2}\int_{M}{\braket{\mathrm{d}_A\delta A\wedge\ast F}}\nonumber\\
&=\frac{1}{4\pi g^2}\int_{M}{\braket{\delta A\wedge\mathrm{d}_A(\ast F)}}+\frac{1}{4\pi g^2}\int_{\partial M}{\braket{\delta A\wedge\ast F}}
\label{eq:II.2}
\end{align}
In case that $M$ has no boundary, the second term vanishes (the case of nontrivial boundary will be discussed below). Thus, in this situation, the variation of the Yang-Mills functional vanishes iff the following Yang-Mills equations are satisfied
\begin{equation}
\mathrm{d}_AF=0\quad\text{and}\quad\mathrm{d}_A(\ast F)=0
\end{equation}
with the former known as the Bianchi identity. In order to construct the corresponding Hamiltonian phase space, we need to perform a 3+1-split of the above action functional. Hence, we assume that $M$ is given by a foliation of the form $M\cong\mathbb{R}\times\Sigma$ with $\Sigma$ a spacelike Cauchy surface. Let $\partial_t$ denote the global timelike vector field induced by the foliation. It follows that $\partial_t=Nn+\vec{N}$ with $n$ a unit normal vector field normal to the time slices $\Sigma_t$ and $\vec{N}$ the shift vector field tangent to the foliation.\\
In analogy to covariant Maxwell theory, we define the Yang-Mills vector potential $\mathbf{A}$ as the spatial part of the YM-gauge field such that $\mathbf{A}=i_{\Sigma}^{*}A$ with $i_{\Sigma}: \Sigma\hookrightarrow M$ the embedding of $\Sigma$ in $M$. Likewise, the YM-electric and magnetic fields (or rather two-forms) are defined via $\mathbf{E}=i_{\Sigma}^{*}(\ast F)$ and $\mathbf{B}=i_{\Sigma}^{*}F$, respectively. In terms of these quantities, it can be shown that the 3+1-split of the action functional (\ref{eq:II.1}) then takes the form   
\begin{align}
S_{\mathrm{YM}}[\mathbf{A},\mathbf{E}]=-\frac{1}{4\pi g^2}\int_{\mathbb{R}}\mathrm{d}t\,\int_{\Sigma_t}\langle\dot{\mathbf{A}}\wedge\mathbf{E}+A(\partial_t)\mathrm{d}_{\mathbf{A}}\mathbf{E}&-\left(i_{\vec{N}}\mathbf{B}\right)\wedge\mathbf{E}-\mathrm{d}(A(\partial_t)\mathbf{E})\rangle\label{eq:C19}\\
&+\frac{N}{2}\braket{\mathbf{E}\wedge\ast\mathbf{E}+\mathbf{B}\wedge\ast\mathbf{B}}\nonumber
\end{align}
where $\dot{\mathbf{A}}=L_{\partial_t}\mathbf{A}$. Thus, Yang-Mills theory is described by a constrained Hamiltonian phase space with canonically conjugate variables $(\mathbf{A},\mathbf{E})$. \\
From mow on, let us assume that $\braket{\cdot,\cdot}$ is minus the Killing metric $B(X,Y):=\mathrm{tr}(\mathrm{ad}(X)\circ\mathrm{ad}(Y))$ of the semisimple compact Lie group $G$. Then, there is a basis $\{\tau_I\}$ of $\mathfrak{g}$ such that $\braket{\tau_I,\tau_J}=\delta_{IJ}$. In this basis, the Poisson bracket reads
\begin{equation}
\{\mathbf{A}_a^I(x),\mathbf{E}^b_J(y)\}=-4\pi g^2\delta_a^b\delta^I_J\delta^{(3)}(x,y)
\label{eq:3.2.24}
\end{equation}
where $\mathbf{E}^{a}_I:=\frac{1}{2}\tensor{\epsilon}{^{abc}}\mathbf{E}_{bc\,I}$ is the corresponding vector density. Thus, the symplectic structure on the phase space $(\mathbf{A},\mathbf{E})$ of the Hamiltonian Yang-Mills theory is given by 
\begin{equation}
\Omega_{\mathrm{YM}}(\delta_1,\delta_2)=-\frac{1}{4\pi g^2}\int_{\Sigma}{\braket{\delta_1\mathbf{A}\wedge\delta_2\mathbf{E}-\delta_2\mathbf{A}\wedge\delta_1\mathbf{E}}}
\label{eq:3.2.25}
\end{equation}
If $M$ has a boundary such as an inner boundary given by an isolated horizon $(H,[l])$, the boundary term in (\ref{eq:II.2}) does not vanish. Hence, in this case one eventually needs to add an additional term to the action functional in order to recover the field equations. However, as discussed in \cite{Ashtekar:1999yj}, this problem can be circumvented by appropriately restricting the phase space. In fact, since $H$ is topologically $\mathbb{R}\times S^2$ and $\ast F=\mathbf{E}+i_n(\ast F)\wedge n^{\flat}$ with $n^{\flat}:=g(n,\cdot)$, it follows that the boundary term in (\ref{eq:II.2}) becomes proportional to
\begin{equation}
\int_{\mathbb{R}}{\mathrm{d}v\int_{S^2_v}{\braket{\delta(A(l))\ast\! F}}}
\label{eq:3.2.26}
\end{equation}
where it was used that $\partial_t=l$ as well as $i_l(\ast F)=0$ on the horizon (\cite{Ashtekar:1999yj}, see also (\ref{eq:3.4.14}) below). One then imposes a gauge fixing by requiring that $A(l)\mathrm{vol}_{S^2}$\footnote{here $\mathrm{vol}_{S^2}$ denotes the volume form on $S^2$ induced by the pullback of the metric $g$.} is proportional to $\ast F$ on $H$. Hence, by additionally restricting the Yang-Mills phase space of histories by fixing the YM-potential $\Phi^{\mathrm{YM}}:=\|A(l)\|$ to a specific universal constant, it follows that the boundary term (\ref{eq:3.2.26}) in the variation of the action functional indeed vanishes.\\
%

\subsection{Isolated horizon boundary conditions in the presence of YM fields}
\label{suse_bc}
The bulk and horizon degrees of freedom are not independent of each other but need to obey certain boundary conditions enforced by the properties of isolated horizons. These classical constraints are then mapped to corresponding quantum boundary conditions in the quantum theory. In the following, we want to derive constraints for the Yang-Mills degrees of freedom generalizing the results found in \cite{Ashtekar:1999wa} in the $\mathrm{U}(1)$ case by working also in the spinor bundle. For type I IH, these results coincide with those derived in \cite{Corichi:2000dm} for non-Abelian gauge theories. We need to compute the energy-momentum tensor $\tensor{T}{_{\mu\nu}}$ which is given by the variation of the action functional (\ref{eq:II.1}) w.r.t. the metric $g_{\mu\nu}$. In doing so, we find    
\begin{equation}
\tensor{T}{_{\mu\nu}}=\frac{2}{\sqrt{-g}}\frac{\delta S_{\mathrm{YM}}}{\delta g^{\mu\nu}}=\frac{1}{4\pi}\left(\braket{\tensor{F}{_{\mu\rho}},\tensor{F}{_{\nu}^{\rho}}}-\frac{1}{4}\tensor{g}{_{\mu\nu}}\braket{\tensor*{F}{_{\alpha\beta}},\tensor*{F}{^{\alpha\beta}}}\right)
\label{eq:C01}
\end{equation}
Mapping this to the spinor bundle, this yields
\begin{equation}
\tensor{T}{_{AA'BB'}}=\frac{1}{4\pi}\left(\tensor{F}{_{AA'CC'}^I}\tensor{F}{_{BB'}^{CC'}^I}-\frac{1}{4}\tensor{\epsilon}{_{AB}}\tensor{\bar{\epsilon}}{_{A'B'}}\tensor{F}{_{CC'DD'}^I}\tensor*{F}{^{CC'DD'}^I}\right)
\label{eq:3.4.1}
\end{equation}
Since the field strength tensor is totally antisymmetric one can decompose it in the spinor bundle in the following way 
\begin{equation}
\tensor{F}{_{AA'BB'}^I}=\tensor{\phi}{_{AB}^I}\tensor{\bar{\epsilon}}{_{A'B'}}+\tensor{\bar{\phi}}{_{A'B'}^I}\tensor{\epsilon}{_{AB}}
\label{eq:3.4.2}
\end{equation}
where $\tensor{\phi}{_{AB}^I}=\tensor{\phi}{_{(AB)}^I}$ and $\tensor{\bar{\phi}}{_{A'B'}^I}=\tensor{\bar{\phi}}{_{(A'B')}^I}$ are both symmetric. By inserting this into the energy-momentum tensor, we find for the first term
\begin{align}
\tensor{F}{_{AA'CC'}^I}\tensor{F}{_{BB'}^{CC'}^I}&=\left(\tensor{\phi}{_{AC}^I}\tensor{\bar{\epsilon}}{_{A'C'}}+\tensor{\bar{\phi}}{_{A'C'}^I}\tensor{\epsilon}{_{AC}}\right)\left(\tensor{\phi}{_B^C^I}\tensor{\delta}{_{B'}^{C'}}+\tensor{\bar{\phi}}{_{B'}^{C'}^I}\tensor{\delta}{_B^C}\right)\nonumber\\
&=\tensor{\phi}{_{AC}^I}\tensor{\phi}{_B^C^I}\tensor{\bar{\epsilon}}{_{A'B'}}+\tensor{\bar{\phi}}{_{B'}^{C'}^I}\tensor{\bar{\phi}}{_{A'C'}^I}\tensor{\epsilon}{_{AB}}-2\tensor{\bar{\phi}}{_{A'B'}^I}\tensor{\phi}{_{AB}^I}
\label{eq:3.4.3}
\end{align}
The first and second term on the right hand side can be further evaluated. For this let us define the tensor $\tensor{C}{_{AB}}=\tensor{\epsilon}{^{CD}}\tensor{\phi}{_{AC}^I}\tensor{\phi}{_{BD}^I}$. By interchanging the tensor indices we then find
\begin{align}
\tensor{C}{_{AB}}=\tensor{\epsilon}{^{CD}}\tensor{\phi}{_{AC}^I}\tensor{\phi}{_{BD}^I}=-\tensor{\epsilon}{^{DC}}\tensor{\phi}{_{AC}^I}\tensor{\phi}{_{BD}^I}=-\tensor{\phi}{_{BD}^I}\tensor{\phi}{_A^D^I}=-\tensor{C}{_{BA}}
\label{eq:3.4.4}
\end{align}
telling us that $\tensor{C}{_{AB}}$ is antisymmetric. Hence, there exists a function $\alpha$ such that $\tensor{C}{_{AB}}=\alpha\tensor{\epsilon}{_{AB}}$. Taking the trace of both sides then yields $\tensor{\epsilon}{^{AB}}\tensor{C}{_{AB}}=2\alpha$ which gives us for $\tensor{C}{_{AB}}$ the simple form 
\begin{equation}
\tensor{C}{_{AB}}=\frac{1}{2}\tensor{\phi}{^{AC}^I}\tensor{\phi}{_{AC}^I}\tensor{\epsilon}{_{AB}}
\label{eq:3.4.6}
\end{equation}
Thus the first term in the energy-momentum tensor reads as follows  
\begin{equation}
\tensor{F}{_{AA'CC'}^I}\tensor{F}{_{BB'}^{CC'}^I}=\frac{1}{2}\tensor{\epsilon}{_{AB}}\tensor{\bar{\epsilon}}{_{A'B'}}\left(\tensor{\phi}{^{CD}^I}\tensor{\phi}{_{CD}^I}+\tensor{\bar{\phi}}{^{C'D'}^I}\tensor{\bar{\phi}}{_{C'D'}^I}\right)-2\tensor{\bar{\phi}}{_{A'B'}^I}\tensor{\phi}{_{AB}^I}
\label{eq:3.4.7}
\end{equation}
From this expression we can directly compute the second term in (\ref{eq:3.4.1}). By taking another trace we immediately get
\begin{align}
\tensor{F}{_{CC'DD'}^I}\tensor{F}{^{CC'DD'}^I}&=\frac{1}{2}\tensor{\delta}{_A^A}\tensor{\delta}{_{B'}^{B'}}\left(\tensor{\phi}{^{CD}^I}\tensor{\phi}{_{CD}^I}+\tensor{\bar{\phi}}{^{C'D'}^I}\tensor{\bar{\phi}}{_{C'D'}^I}\right)-2\tensor{\bar{\phi}}{_{A'}^{A'}^I}\tensor{\phi}{_{A}^{A}^I}\nonumber\\
&=2\tensor{\phi}{^{CD}^I}\tensor{\phi}{_{CD}^I}+2\tensor{\bar{\phi}}{^{C'D'}^I}\tensor{\bar{\phi}}{_{C'D'}^I}
\label{eq:3.4.8}
\end{align}
Hence, by inserting (\ref{eq:3.4.7}) and (\ref{eq:3.4.8}) into equation (\ref{eq:3.4.1}) we arrive at a very compact form of the energy-momentum tensor for the Yang-Mills field
\begin{equation}
\tensor{T}{_{AA'BB'}}=-\frac{1}{2\pi}\tensor{\phi}{_{AB}^I}\tensor{\bar{\phi}}{_{A'B'}^I}
\label{eq:3.4.9}
\end{equation}
Let us extend $\braket{\cdot,\cdot}$ to a positive definite sesquiliniear form of the complexified Lie algebra $\mathfrak{g}_{\mathbb{C}}$ by setting 
\begin{equation}
\braket{X\otimes w,Y\otimes z}=\bar{w}z\braket{X,Y}\quad\forall X,Y\in\mathfrak{g},z,w\in\mathbb{C}
\end{equation}
With respect to this metric, (\ref{eq:3.4.9}) can be written as
\begin{equation}
\tensor{T}{_{AA'BB'}}=-\frac{1}{2\pi}\tensor{\phi}{_{AB}^I}\tensor{\bar{\phi}}{_{A'B'}^I}=-\frac{1}{2\pi}\braket{\tensor{\phi}{_{AB}},\tensor{\phi}{_{AB}}}=:-\frac{1}{2\pi}\|\tensor{\phi}{_{AB}}\|^2
\end{equation} 
The boundary conditions for non-expanding horizons restrict the matter degrees of freedom at the horizon. Indeed, as a direct consequence of the matter conditions imposed on general NEHs $(H,[l])$ it follows that the tensor component $\tensor{T}{_{\mu\nu}}l^{\mu}l^{\nu}$ along the future-directed null normal $l$ has to vanish \cite{Ashtekar:2000hw}. If we reformulate this in spinor language, this immediately gives  
\begin{align}
\tensor{T}{_{AA'BB'}}\tensor{l}{^{AA'}}\tensor{l}{^{BB'}}&=-\frac{1}{2\pi}o^Ao^B\tensor{\phi}{_{AB}^I}\bar{o}^{A'}\bar{o}^{B'}\tensor{\bar{\phi}}{_{A'B'}^I}\nonumber\\
&=-\frac{1}{2\pi}\|o^Ao^B\tensor{\phi}{_{AB}}\|^2=0
\label{eq:3.4.10}
\end{align}
that is, due to the positive definiteness of $\|\cdot\|$,
\begin{equation}
o^Ao^B\tensor{\phi}{_{AB}^I}=0
\label{eq:3.4.11}
\end{equation}
For $G=\mathrm{U}(1)$ this precisely coincides with the result obtained in \cite{Ashtekar:1999wa}. Thus, at the horizon the $i_Ai_B$ component of $\tensor{\phi}{_{AB}}$ has to vanish such that
\begin{equation}
\tensor{\phi}{_{AB}^I}=\tensor{\phi}{_{(AB)}^I}=-2\phi_1^I\,\tensor{i}{_{(A}}\tensor{o}{_{B)}}+\phi_2^I\,\tensor{o}{_{A}}\tensor{o}{_{B}}
\label{eq:3.4.12}
\end{equation}
where $\phi_1$ and $\phi_2$ are two Newman-Penrose coefficients. Let us convert these results back to the tangent bundle using the soldering form (\ref{eq:B7}). The field strength then takes the form 
\begin{align}
\tensor{F}{_{\mu\nu}^I}&=\tensor*{\sigma}{_{[\mu}^{AA'}}\tensor*{\sigma}{_{\nu]}^{BB'}}\tensor{F}{_{AA'BB'}^I}\nonumber\\
&=\tensor{\phi}{_{AB}^I}\tensor*{\sigma}{_{[\mu}^{AA'}}\tensor*{\sigma}{_{\nu]}^{BB'}}\tensor{\bar{\epsilon}}{_{A'B'}}+\tensor{\bar{\phi}}{_{A'B'}^I}\tensor*{\sigma}{_{[\mu}^{AA'}}\tensor*{\sigma}{_{\nu]}^{BB'}}\tensor{\epsilon}{_{AB}}\nonumber\\
&=\tensor{\phi}{_{AB}^I}\tensor*{\Sigma}{_{\mu\nu}^{AB}}+\tensor{\bar{\phi}}{_{A'B'}^I}\tensor*{\bar{\Sigma}}{_{\mu\nu}^{A'B'}}
\label{eq:3.4.13}
\end{align}
Together with (\ref{eq:3.4.12}) and the formula (\ref{eq:2.2.3}) for $\Sigma^{AB}$ this yields
\begin{align}
F^I&=\left(-2\phi_1^I\,\tensor{i}{_{(A}}\tensor{o}{_{B)}}+\phi_2^I\,\tensor{o}{_{A}}\tensor{o}{_{B}}\right)\tensor{\Sigma}{^{AB}}+\left(-2\bar{\phi}_1^I\,\tensor{\bar{i}}{_{(A'}}\tensor{\bar{o}}{_{B')}}+\bar{\phi}_2^I\,\tensor{\bar{o}}{_{A'}}\tensor{\bar{o}}{_{B'}}\right)\tensor{\bar{\Sigma}}{^{AB}}\nonumber\\
&=-\phi_1^I(m\wedge\bar{m}-l\wedge k)-\phi_2^I\,l\wedge m+\bar{\phi}_1^I(m\wedge\bar{m}+l\wedge k)-\bar{\phi}_2^I\,l\wedge\bar{m}\nonumber\\
&=-2\mathrm{Im}(\phi_1^I)\,im\wedge\bar{m}+2\mathrm{Re}(\phi_1^I)\,l\wedge k-\phi_2^I\,l\wedge m-\bar{\phi}_2^I\,l\wedge\bar{m}
\label{eq:3.4.14}
\end{align}
We are now ready to derive the horizon boundary conditions for the phase space variables of the Hamiltonian Yang-Mills theory. Recall that the Yang-Mills magnetic $\mathbf{B}$-field is defined as the spatial part of the field strength. Thus by pulling back formula (\ref{eq:3.4.14}) to the 2-sphere cross-section of the horizon we find that
\begin{equation}
\underset{\raisebox{1pt}{$\Leftarrow$}}{\mathbf{B}}^I=\underset{\raisebox{1pt}{$\Leftarrow$}}{F}^I=-2\mathrm{Im}(\phi_1^I)\,\mathrm{vol}_{S^2}
\label{eq:3.4.15}
\end{equation}
where we used that the only remaining term under pull-back is the one proportional to the volume form of the 2-sphere $\mathrm{vol}_{S^2}=im\wedge\bar{m}$.\\
Likewise, the Yang-Mills electric $\mathbf{E}$-field is given by the spatial projection of the dual field strength tensor. Taking the Hodge-dual of (\ref{eq:3.4.14}) and noticing that $\ast(l\wedge k)=\ast(e^0\wedge e^3)=-e^1\wedge e^2=-\mathrm{vol}_{S^2}$ the pullback to $S^2$ yields
\begin{equation}
\underset{\raisebox{1pt}{$\Leftarrow$}}{\mathbf{E}}^I=\underset{\raisebox{1pt}{$\Leftarrow$}}{\ast F}^I=-2\mathrm{Re}(\phi_1^I)\,\mathrm{vol}_{S^2}
\label{eq:3.4.16}
\end{equation}
Equation (\ref{eq:3.4.15}) and (\ref{eq:3.4.16}) describe the coupling between the Yang-Mills and horizon degrees of freedom which hold for all NEHs. We will need them later when we want to go over to the quantum description of general distorted black holes.\\
Let us specialize our formulas to the case of type I isolated horizons. The additional spherical symmetry assumption will then lead to a further restriction of the matter degrees of freedom at the horizon. Indeed, since the matrix element $\tensor{T}{_{\mu\nu}}l^{\mu}k^{\nu}$ is spherically symmetric \cite{Ashtekar:2000hw}, this implies that
\begin{align}
\tensor{T}{_{AA'BB'}}l^{AA'}k^{BB'}&=-\frac{1}{2\pi}\tensor{\phi}{_{AB}^I}\tensor{\bar{\phi}}{_{A'B'}^I}i^Ao^B\bar{i}^{A'}\bar{o}^{B'}=-\frac{1}{2\pi}\|i^Ao^B\tensor{\phi}{_{AB}}\|^2\nonumber\\
&=-\frac{1}{2\pi}\|\phi_1\|^2
\label{eq:3.4.17}
\end{align}   
also has to be spherically symmetric. Using this, let us compute the electric charge $Q_H$ of the black hole. Following \cite{Ashtekar:2000hw,Corichi:2000dm}, in analogy to Gauss' law in classical electrodynamics, we define the total charge as the electric flux through the closed 2-sphere
\begin{equation}
Q_{H}:=\frac{1}{4\pi}\int_{S^2}{\|\underset{\raisebox{1pt}{$\Leftarrow$}}{\mathbf{E}}\|}
\end{equation}
And likewise for the magnetic charge
\begin{equation}
P_{H}=\frac{1}{4\pi}\int_{S^2}{\|\underset{\raisebox{1pt}{$\Leftarrow$}}{\mathbf{B}}\|}
\end{equation}
Here, $\|\mathbf{E}\|$ is a real-valued two-form on $M$ defined via $\|\mathbf{E}\|(X,Y):=\|\mathbf{E}(X,Y)\|=\sqrt{\braket{\mathbf{E}(X,Y),\mathbf{E}(X,Y)}}$ for any $X,Y\in\mathfrak{g}$ and likewise for $\|\mathbf{B}\|$. In case of Type I IH, we saw by (\ref{eq:3.4.14}) that the norm of the Penrose coefficient $\phi_1$ becomes spherically symmetric. Hence, we find   
\begin{equation}
Q_{H}=\frac{1}{4\pi}\int_{S^2}{\|\underset{\raisebox{1pt}{$\Leftarrow$}}{\mathbf{E}}\|}=\frac{\|\mathrm{Re}(\phi_1)\|}{2\pi}\int_{S^2}{\mathrm{vol}_{S^2}}=\frac{a_{H}}{2\pi}\|\mathrm{Re}(\phi_1)\|
\end{equation}
For the magnetic charges we get
\begin{equation}
P_{H}=\frac{1}{4\pi}\int_{S^2}{\|\underset{\raisebox{1pt}{$\Leftarrow$}}{\mathbf{B}}\|}=\frac{\|\mathrm{Im}(\phi_1)\|}{2\pi}\int_{S^2}{\mathrm{vol}_{S^2}}=\frac{a_H}{2\pi}\|\mathrm{Im}(\phi_1)\|
\end{equation}
and thus \cite{Corichi:2000dm}
\begin{equation}
\|\mathrm{Re}(\phi_1)\|=\frac{2\pi}{a_H}Q_H\quad\mathrm{and}\quad\|\mathrm{Im}(\phi_1)\|=\frac{2\pi}{a_H}P_H
\label{eq:II.32}
\end{equation}
For $G=\mathrm{U}(1)$ the norm in the definition of $Q_H$ and $P_H$ can be dropped. In this case, one simply sets
\begin{equation}
Q_H=-\frac{1}{4\pi}\int_{S^2}{\underset{\raisebox{1pt}{$\Leftarrow$}}{\mathbf{E}}}
\label{eq:II.33}
\end{equation}
and likewise for $P_H$. For spherically symmetric isolated horizons, a similar argument as above then yields \cite{Ashtekar:1999wa}
\begin{equation}
\phi=\frac{2\pi}{a_H}(Q_H+iP_H)
\label{eq:II.34}
\end{equation}

\subsection{Phase space}
We want to discuss the quantum theory of generic non-rotating distorted black holes coupled to a Yang-Mills field. Therefore, it is convenient to apply the machinery of the full $\mathrm{SU}(2)$ approach as developed in \cite{Perez:2010pq}, since it treats distorted black holes on an equal footing right from the beginning.\\
Here, the horizon degrees of freedom of the classical phase space $\Gamma$ of the Hamiltonian Einstein-Yang-Mills theory are given by two $\mathrm{SU}(2)$ connections $A_{\sigma_{+}}$ and $A_{\sigma_{-}}$ defined as 
\begin{equation}
A^i_{\sigma_{\pm}}=\Gamma^i+\sqrt{\frac{2\pi}{a_H}}\sigma_{\pm}e^i
\end{equation}
parametrized by two real numbers $\sigma_{\pm}$, where $\Gamma^i$ is the spin-connection and $\{e^i\}$ a spatial triad. The symplectic structure on $\Gamma$ then takes the form
\begin{equation}
\Omega^{\Gamma}(\delta_1,\delta_2)=\Omega^{(A,E)}_{\mathrm{grav}}(\delta_1,\delta_2)+\Omega^{(\mathbf{A},\mathbf{E})}_{\mathrm{YM}}(\delta_1,\delta_2)+\Omega^{A_{\sigma_{+}}}_{\mathrm{CS}}(\delta_1,\delta_2)+\Omega^{A_{\sigma_{-}}}_{\mathrm{CS}}(\delta_1,\delta_2)
\label{eq:3.6.1.1}
\end{equation}
with $\Omega^{(\mathbf{A},\mathbf{E})}_{\mathrm{YM}}$ given by (\ref{eq:3.2.25}) and $\Omega^{(A,E)}_{\mathrm{grav}}$ the symplectic structure of the pure gravitational degrees of freedom in the bulk expressed in terms of the Ashtekar-Barbero variables $A_a^i$ and $E^a_i$, that is,
\begin{equation}
\Omega^{(A,E)}_{\mathrm{grav}}(\delta_1,\delta_2)=\frac{2}{\beta\kappa}\int_{\Sigma}{\mathrm{d}^3x\,\delta_1 A_a^i\delta_2E^a_i-\delta_2 A_a^i\delta_1E^a_i}
\end{equation}
Furthermore, in the $\mathrm{SU}(2)$ framework, the boundary symplectic structure $\Omega^{A_{\sigma_{\pm}}}_{\mathrm{CS}}$ is given by the symplectic structures of two $\mathrm{SU}(2)$ Chern-Simons gauge theories which read 
\begin{equation}
\kappa\beta\Omega^{A_{\sigma_{\pm}}}_{\mathrm{CS}}(\delta_1,\delta_2)=\mp\frac{a_H}{\pi(\sigma_{-}^2-\sigma_{+}^2)}\int_{S^2}{\delta_1A_{\sigma_{\pm}}^i\wedge\delta_2A_{\sigma_{\pm} i}}
\label{eq:3.6.1.2}
\end{equation}
The bulk and horizon degrees of freedom obey certain boundary conditions at the black hole horizon. For the Yang-Mills magnetic and electric field, these are given by the constraints (\ref{eq:3.4.15}) and (\ref{eq:3.4.16}), respectively, which were derived for generic NEHs. Concerning the gravitational degrees of freedom, we first need to re-express the well-known boundary condition   
\begin{equation}
\underset{\raisebox{1pt}{$\Longleftarrow$}}{F(A^+)}=\left(\Psi_2-\Phi_{11}-\frac{R}{24}\right)\underset{\raisebox{1pt}{$\Leftarrow$}}{\Sigma}^{+}
\label{eq:3.6.1.3}
\end{equation}
for the self-dual Ashtekar connection $A^{+}=\Gamma+iK$ in terms of the connections $A_{\sigma_{+}}$ and $A_{\sigma_{-}}$. Therefore, if we choose a basis $\{\tau_i\}$ of $\mathfrak{su}(2)$ with $[\tau_i,\tau_j]=\tensor{\epsilon}{^k_{ij}}\tau_k$, we notice that the self-dual part of $\Sigma^{IJ}:=e^I\wedge e^J$ can be written as $\Sigma^{+i}=\tensor{\epsilon}{^i_{jk}}\Sigma^{jk}+2i\Sigma^{0i}$. Hence, for the pullback to the 2-sphere cross-section of the horizon, we find that \cite{Engle:2010kt}
\begin{equation}
\underset{\raisebox{1pt}{$\Leftarrow$}}{\Sigma}^{+i}=\tensor{\epsilon}{^i_{jk}}e^{j}\wedge e^{k}=:\Sigma^{i}
\label{eq:3.6.1.4}
\end{equation}
For convenience, let us suppress the arrow symbol in the following. If not otherwise stated the pullback to $S^2$ will always be assumed. In the basis $\{\tau_i\}$ as chosen above, the curvature of a connection $A$ can be written as     
\begin{equation}
F(A)^i=\mathrm{d}A^i+\frac{1}{2}\tensor{\epsilon}{^i_{kl}}A^k\wedge A^l
\label{eq:3.6.1.5}
\end{equation}
Hence, for the self-dual Ashtekar connection $A^+$ we compute
\begin{align}
F(A^+)^i&=\mathrm{d}\Gamma^i+i\mathrm{d}K^i+\frac{1}{2}\tensor{\epsilon}{^i_{kl}}(\Gamma^k+iK^k)\wedge(\Gamma^l+iK^l)\nonumber\\
&=\mathrm{d}\Gamma^i+\frac{1}{2}\tensor{\epsilon}{^i_{kl}}\Gamma^k\wedge\Gamma^l-\frac{1}{2}\tensor{\epsilon}{^i_{kl}}K^k\wedge K^l+i\mathrm{d}K^i+i\frac{1}{2}\tensor{\epsilon}{^i_{kl}}(\Gamma^k\wedge K^l+K^k\wedge\Gamma^l)\nonumber\\
&=F(\Gamma)^i-\frac{c}{2}\Sigma^i+i(\mathrm{d}K^i+\tensor{\epsilon}{^i_{kl}}\Gamma^k\wedge K^l)\nonumber\\
&=F(\Gamma)^i-\frac{c}{2}\Sigma^i+i\mathrm{d}_{\Gamma}K^i
\label{eq:3.6.1.6}
\end{align}
where we inserted the definition of the exterior covariant derivative $\mathrm{d}_{\Gamma}$ in the adjoint representation induced by the spin-connection $\Gamma$. Moreover, in the third line, we used the identity \cite{Engle:2010kt}
\begin{equation}
\tensor{\epsilon}{^i_{kl}}K^k\wedge K^l=c\Sigma^i
\label{eq:3.6.1.7}
\end{equation}
with $c:\,H\rightarrow\mathbb{R}$ an extrinsic curvature scalar. Thus, since $\mathrm{Im}\,\Psi_2=0$ for static IH and $\Phi_{11}$ and $R$ are real by definition, we see that relation (\ref{eq:3.6.1.3}) implies that the curvature of the self-dual connection is purely real, i.e. $\mathrm{Im}\,F(A^+)=0$. Therefore, from (\ref{eq:3.6.1.6}) we deduce that at the horizon one has \cite{Engle:2010kt}
\begin{equation}
\mathrm{d}_{\Gamma}K^i=0
\label{eq:3.6.1.8}
\end{equation}
Reinserting this into (\ref{eq:3.6.1.6}), we thus find  
\begin{equation}
F(\Gamma)^i=F(A^+)^i+\frac{c}{2}\Sigma^i
\label{eq:3.6.1.9}
\end{equation}
With this relation, we are ready to compute the curvature of the connection $A_{\sigma_{\pm}}$. Indeed, using (\ref{eq:3.6.1.5}), we get for $A_{\sigma}:=\Gamma+\sigma e$  
\begin{align}
F(A_{\sigma})^i&=\mathrm{d}\Gamma^i+\sigma\mathrm{d}e^i+\frac{1}{2}\tensor{\epsilon}{^i_{kl}}(\Gamma^k+\sigma e^k)\wedge(\Gamma^l+\sigma e^l)\nonumber\\
&=\mathrm{d}\Gamma^i+\frac{1}{2}\tensor{\epsilon}{^i_{kl}}\Gamma^k\wedge\Gamma^l+\sigma(\mathrm{d}e^i+\tensor{\epsilon}{^i_{kl}}\Gamma^k\wedge e^l)+\sigma^2\frac{1}{2}\tensor{\epsilon}{^i_{kl}}e^k\wedge e^l\nonumber\\
&=F(\Gamma)^i+\frac{\sigma^2}{2}\Sigma^i+\mathrm{d}_{\Gamma}e^i\nonumber\\
&=F(A^+)^i+\frac{1}{2}(\sigma^2+c)\Sigma^i
\label{eq:3.6.1.10}
\end{align}
where, from the third to the last line, we used that $\mathrm{d}_{\Gamma}e=0$, since $\Gamma$ is torsion-free. Together with (\ref{eq:3.6.1.3}), this yields the desired formula
\begin{equation}
F(A_{\sigma_{\pm}})^i=\left(\Psi_2-\Phi_{11}-\frac{R}{24}\right)\Sigma^i+\left(\frac{\pi}{a_H}\sigma_{\pm}^2+\frac{c}{2}\right)\Sigma^i
\label{eq:3.6.1.11}
\end{equation}
describing the coupling between the horizon and the bulk degrees of freedom. This is a generalization of the identity found in \cite{Perez:2010pq} which now also holds for matter fields at the horizon. Let us specialize this formula for the case of YM fields coupled to gravity. Therefore, we first observe that taking the trace of the Einstein field equations (with $\Lambda=0$) yields 
\begin{equation}
8\pi T=R-\frac{1}{2}Rg^{\mu\nu}g_{\mu\nu}=R-2R=-R
\label{eq:3.6.1.12}
\end{equation}
Furthermore, using the energy-momentum tensor (\ref{eq:C01}) for the Yang-Mills field, we find
\begin{equation}
T=g^{\mu\nu}T_{\mu\nu}=\frac{1}{4\pi}\left(\braket{F_{\mu\rho},F^{\mu\rho}}-\braket{F_{\alpha\beta}F^{\alpha\beta}}\right)=0
\label{eq:3.6.1.13}
\end{equation}
which, due to (\ref{eq:3.6.1.12}), then tells us that $R=0$, that is, the scalar curvature vanishes in the Einstein-Yang-Mills theory. Next, we notice that $\Phi_{11}$ can completely be expressed in terms of the Yang-Mills degrees of freedom. In fact, since by definition, $\Phi_{11}=2\pi T_{\mu\nu}(l^{\mu}k^{\nu}+m^{\mu}\bar{m}^{\nu})$, we see that, using (\ref{eq:3.4.9}) and (\ref{eq:3.4.12}),
\begin{align}
\Phi_{11}&=2\pi T_{\mu\nu}(l^{\mu}k^{\nu}+m^{\mu}\bar{m}^{\nu})\nonumber\\
&=2\pi T_{AA'BB'}(l^{AA'}k^{BB'}+m^{AA'}\bar{m}^{BB'})\nonumber\\
&=\phi_{AB}^I\bar{\phi}_{A'B'}^I(o^A\bar{o}^{A'}i^B\bar{i}^{B'}+o^A\bar{i}^{A'}i^B\bar{o}^{B'})\nonumber\\
&=2\|o^Ai^B\phi_{AB}\|^2=8\|\phi_1i_{(A}o_{B)}o^Ai^B\|^2\nonumber\\
&=2\|\phi_1\|^2
\label{eq:3.6.1.14}
\end{align}
Hence, we conclude that the horizon boundary conditions of the symplectic phase space $(\Gamma,\Omega^{\Gamma})$ are given by (\ref{eq:3.4.15}) and (\ref{eq:3.4.16}) for the Yang-Mills fields as well as the constraints
\begin{equation}
F(A_{\sigma_{\pm}})^i=\left(\Psi_2-2\|\phi_{1}\|^2\right)\Sigma^i+\left(\frac{\pi}{a_H}\sigma_{\pm}^2+\frac{c}{2}\right)\Sigma^i
\label{eq:3.6.1.15}
\end{equation}
connecting the gravitational field in the bulk with the horizon degrees of freedom which, due to (\ref{eq:3.6.1.15}), are described by two $\mathrm{SU}(2)$ Chern-Simons theories on the 2-sphere with punctures.

\section{Quantized charged type III horizon} 
\label{se_typeIII}
\subsection{Quantization of Yang-Mills theory}
The symplectic phase space of the Yang-Mills degrees of freedom is quantized by using techniques of loop quantum gravity. This results in a background independent quantum theory, see 
\cite{Thiemann:1997rt} for the general case, and \cite{Corichi:1997us} for U(1). This formalism is suitable for black holes with vanishing magnetic charge. We will elaborate on the standard treatment by presenting the quantization of an observable related to the electric flux, and horizon electric charge $Q_H$, in detail. 
 
The case of non-vanishing magnetic charge $P_H$ is more complicated, as the standard holonomy observables used in LQG are not suitable to detect magnetic flux. Thus this case needs a substantial generalization of the formalism of \cite{Thiemann:1997rt,Corichi:1997us} as far as we see. In the case of structure group U(1), we have found a relatively straightforward way to obtain a  generalization that is sufficient to  allow for operators corresponding to exponentiated magnetic fluxes, and for an operator corresponding to the magnetic horizon charge $P_H$ in particular. 
We will start with the case of $P_H=0$, and subsequently explain the generalization in the case of U(1). 
\subsubsection{Vanishing magnetic horizon charge}
The YM-vector potential $\mathbf{A}_a^I$ is smeared over one-dimensional paths $p\subset\Sigma$ assigning to it the corresponding holonomy
\begin{equation}
h_p(\mathbf{A})=\mathcal{P}\exp\left(\int_p{\mathbf{A}}\right)
\label{eq:III.1}
\end{equation}
which defines an element of $\bar{\mathcal{A}}_G=\mathrm{Hom}(\mathcal{P},G)$ called the space of generalized holonomies on the path groupoid $\mathcal{P}$. The YM-electric fields are smeared over two-dimensional oriented surfaces $S$ embedded in $\Sigma$ yielding the electric fluxes
\begin{equation}
\mathbf{E}_n(S)=\frac{1}{g}\int_S{n^I\mathbf{E}_I}
\label{eq:III.2}
\end{equation}
where $n^I$ is some Lie algebra-valued smearing function defined on $S$. The holonomies and fluxes then give rise to the so-called holonomy-flux Poisson algebra $\mathfrak{P}$ given by the space of cylindrical functions $\mathrm{Cyl}$ and vector fields $X_n(S)\in V(\mathrm{Cyl})$ thereon defined via (see e.g. \cite{Thiemann:2007zz} and references therein)
\begin{equation}
X_n(S)f_{\gamma}:=\{f_{\gamma},\mathbf{E}_n(S)\}=-2\pi g\sum_{e\cap\gamma}{\epsilon(e,S)n(b(e))_IR^I_e}f_{\gamma}
\label{eq:III.3}
\end{equation}
for any cylindrical function $f_{\gamma}$ associated to a graph $\gamma$ in $\Sigma$ where $R^I$ is the right-invariant vector field of $G$. For the quantum theory, we choose as the representation of the holonomy-flux algebra the Ashtekar-Lewandowski representation. Accordingly, the Hilbert space of the quantized Yang-Mills theory is given by $\mathcal{H}_{\mathrm{YM}}=L^2(\bar{\mathcal{A}}_{G},\mathrm{d}\mu_{\mathrm{AL}}^G)$ with $\mu_{\mathrm{AL}}^G$ the Ashtekar-Lewandowski measure induced by the unique Haar-measure on the compact Lie group $G$. On this Hilbert space, the cylindrical functions are represented by multiplication operators and the electric fluxes by derivations, that is,
\begin{equation}
\widehat{f}_{\gamma}:=f_{\gamma}\quad\quad\text{and}\quad\quad\widehat{\mathbf{E}}_n(S)=-iX_n(S)
\label{eq:III.4}
\end{equation}
An orthonormal basis of the Hilbert space $\mathcal{H}_{\mathrm{YM}}$ is given by charge-network states $f_{\gamma,\underline{\lambda}}\equiv\ket{\gamma,\underline{\lambda}}$ with edges $e\in E(\gamma)$ of the graph labeled by highest weights $\lambda_e$ classifying the irreducible representations of the semisimple Lie group $G$.\\
As seen in section \ref{suse_bc}, in analogy to classical electrodynamics, it is possible to associate electric charge to closed two-dimensional surfaces. This leads to the notion of an electric charge operator in the quantum theory, where quantization can be performed in complete analogy to the area operator in quantum geometry. Therefore, we first need to regularize the classical quantity. Hence, suppose that the (not necessarily closed) surface $S$ is contained within a single chart $(U,\phi_U)$. Consider a partition $\mathcal{U}_{\epsilon}=\{U_i\}$ of $U$ of fineness $\epsilon>0$ such that the $S_{U_i}$ defined via $S_{U_i}:=\phi_{U}(U_i)$ cover $S$. We set
\begin{equation}
Q_{\epsilon}(S)=\frac{1}{4\pi}\sum_{U\in\mathcal{U}_{\epsilon}}{\|\mathbf{E}(S_U)\|}=\frac{1}{4\pi}\sum_{U\in\mathcal{U}_{\epsilon}}{\sqrt{\delta^{IJ}\mathbf{E}_I(S_U)\mathbf{E}_J(S_U)}}
\end{equation} 
In the limit $\epsilon\rightarrow 0$, we then have $Q(S)=\lim_{\epsilon\rightarrow 0}Q_{\epsilon}(S)$. The quantization of $Q_{\epsilon}(S)$ is now straightforward. If we replace the electric fluxes by the corresponding operators (\ref{eq:III.4}) in the quantum theory, we get   
\begin{equation}
\widehat{Q}(S)=\lim_{\epsilon\rightarrow 0}\widehat{Q}_{\epsilon}(S), \qquad 
\widehat{Q}_{\epsilon}(S)=\frac{1}{4\pi}\sum_{U\in\mathcal{U}_{\epsilon}}{\sqrt{\delta^{IJ}\widehat{\mathbf{E}}_I(S_U)\widehat{\mathbf{E}}_J(S_U)}}
\label{eq:YM_Q}
\end{equation}
By inserting formula (\ref{eq:III.3}), we find
\begin{align}
\delta^{IJ}\widehat{\mathbf{E}}_I(S_U)\widehat{\mathbf{E}}_J(S_U)&=-4\pi^2g^2\left(\sum_{e\cap S_U\neq\emptyset}{\epsilon(e,S_U)R^I_e}\right)^2\nonumber\\
&=-4\pi^2g^2\left(R_{\mathrm{in}}^I-R_{\mathrm{out}}^I\right)^2=-4\pi^2g^2\left(\left(R_{\mathrm{in}}^I\right)^2+\left(R_{\mathrm{out}}^I\right)^2-2R_{\mathrm{in}}^IR_{\mathrm{out}}^I\right)\nonumber\\
&=-4\pi^2g^2\left(2\left(R_{\mathrm{in}}^I\right)^2+2\left(R_{\mathrm{out}}^I\right)^2-\left(R_{\mathrm{in}}^I+R_{\mathrm{out}}^I\right)^2\right)\nonumber\\
&=:4\pi^2g^2\left(2\Delta_{I}+2\Delta_F-\Delta_{I\cup F}\right)
\end{align}
where $R_{\mathrm{in}}^I=\sum_{e\,\mathrm{ingoing}}{R^I_e}$ and $R_{\mathrm{out}}^I=\sum_{e\,\mathrm{outgoing}}{R^I_e}$ and $\Delta:=-(R^I)^2$ denotes the positive definite Laplace-Beltrami operator of the compact Lie group $G$. In case that $S$ intersects the graph $\gamma$ only in bivalent vertices one has $\Delta_I=\Delta_F$ and $\Delta_{I\cup F}=0$ at any intersection point. Hence, in this case, it follows  
\begin{equation}
\widehat{Q}(S)\ket{\gamma,\underline{\lambda}}=g\sum_{p\in \gamma\cap S}{\sqrt{\Delta_{I(p)}}}\ket{\gamma,\underline{\lambda}}=:g\left(\sum_{p\in \gamma\cap S}{Q(\lambda_p)}\right)\ket{\gamma,\underline{\lambda}}
\end{equation}
Let us again consider the special case $G=\mathrm{U}(1)$. An orthonormal basis of the Hilbert space $\mathcal{H}_{\mathrm{EM}}=L^2(\bar{\mathcal{A}}_{\mathrm{U}(1)},\mathrm{d}\mu_{\mathrm{AL}}^{\mathrm{U}(1)})$ is given by charge-network states $\ket{\gamma,\underline{n}}$ associated to graphs $\gamma$ labeled by integers $n_e\in\mathbb{Z}$ for any edge $e\in E(\gamma)$. The charge operator is then simply given by (\ref{eq:III.4}), i.e. $\widehat{Q}(S)=-\frac{1}{4\pi}\widehat{\mathbf{E}}(S)$, such that
\begin{equation}
\label{eq_QU1}
\widehat{Q}(S)\ket{\gamma,\underline{n}}=g\left(\sum_{e\cap S}{n_e}\right)\ket{\gamma,\underline{n}}
\end{equation} 
where it was used that $\braket{\mathbf{A}|\gamma,\underline{n}}=h_{e_1}(\mathbf{A})^{n_1}\dots h_{e_k}(\mathbf{A})^{n_k}$ for $\gamma=\{e_1,\ldots,e_k\}$ and therefore $R_{e_i}\ket{\gamma,\underline{n}}=in_i\ket{\gamma,\underline{n}}$. That is, the total charge contained in a region bounded by $S$ is an integer multiply of the elementary charge $g$. 

We note that the fact that $\widehat{Q}(S)=-\frac{1}{4\pi}\widehat{\mathbf{E}}(S)$ in the U(1) case has an interesting consequence for the charge network states in the case of the charged black hole. It shows that the flux is really a function on the homotopy classes of surfaces. In particular, any closed genus 0 surface containing the black hole will have the same flux. This means that the charge network state coupling to the CS theory states on the black hole horizon will have to be infinitely extended. We will elaborate on this observation in the following section.
\subsubsection{Nonvanishing magnetic charge for the case of U(1)}
\label{ap_mag}
Electric charge can be measured as electric flux through closed surfaces. As seen in the previous section, the standard formalism in LQG is well equipped to handle this, since electric fluxes through arbitrary surfaces are well defined quantum observables on the kinematic level. By contrast, describing situations  with magnetic charges in the quantum theory runs into difficulties, since magnetic flux is not an observable in the quantum theory. Holonomies are observables, and they are related to exponentiated fluxes by Stokes' theorem, or suitable generalizations in the non-Abelian case, but this relation breaks down precisely when magnetic charges are present. To usefully describe magnetic charges, one would have to start with magnetic fluxes as basic observables in the quantum theory.  We will sketch a realization of this idea for $G=U(1)$. This topic will be more systematically explored and generalized to the non-Abelian case elsewhere \cite{ESS}. 

For U(1), to keep track of the magnetic flux, we have to go over to a new algebra generated by quantities
\begin{equation}
\mathbf{E}(S), \qquad \mathbf{H}(S)
\end{equation}
labeled by (suitably regular) oriented surfaces $S$ in $\Sigma$. The surfaces may be closed, or possess a boundary, and to simplify notation we allow them to consist of multiple disconnected components. The orientation of the boundary is taken to be compatible with that of the surfaces.  

The generators $\mathbf{E}(S), \mathbf{H}(S)$ correspond to electric flux, and exponentiated magnetic flux, respectively, 
\begin{equation}
\mathbf{E}(S) \widehat{=} \frac{1}{g}\int_S \mathbf{E}, \qquad \mathbf{H}(S)  \widehat{=} \exp \int_S d\mathbf{A}.  
\end{equation}
Note that in case $\mathbf{A}$ is smooth and $S$ is compact, $\mathbf{H}(S)$ also corresponds to the holonomy $\mathbf{h}_{\partial S}$ by Stokes' theorem.    
The generators fulfill the relations 
\begin{equation}
\mathbf{E}(S_1)+\mathbf{E}(S_2)= \mathbf{E}(S_1+S_2), \qquad
\mathbf{H}_{S_1}\mathbf{H}_{S_2}=\mathbf{H}_{S_1+S_2}, 
\end{equation}
where the addition is union. Furthermore
\begin{equation}
\mathbf{E}(-S)=-\mathbf{E}(S), \qquad  \mathbf{H}_{-S}=\mathbf{H}_{S}^{-1}
\end{equation}
where inversion is change of orientation, and 
\begin{equation}
\mathbf{E}(S)^*=\mathbf{E}(S), \qquad \mathbf{H}_{S}^{*}=\mathbf{H}_{S}^{-1}. 
\end{equation}
Finally, and most importantly, the canonical commutation relations amount to 
\begin{equation}
[\mathbf{E}(S_1), \mathbf{H}_{S_2}]= g I(S_1, \partial S_2)\mathbf{H}_{S_2},
\label{eq_comm}
\end{equation}
where  $I(S_1, \partial S_2)$ is the signed intersection number of $\partial S_2$ with $S_1$. 
Note that the magnetic fluxes through closed surfaces are in the center of the algebra generated by these objects and relations. 

Now we will construct a representation in which there is a non-trivial magnetic flux. This flux is background data, it can not be changed by the operators of the representation. The construction we use is a variation on that of \cite{Koslowski:2011vn,Campiglia:2013nva,Campiglia:2014hoa}.
Given a classical magnetic field $\mathbf{B}^{(0)}$, the algebra can be represented on the Hilbert space 
\begin{equation}
\label{eq_hilb}
\mathcal{H}_{\mathrm{YM,0}}=L^2(\bar{\mathcal{A}}_{U(1)},\mathrm{d}\mu_{\mathrm{AL}}^{U(1)})
\end{equation}
as follows:
\begin{equation}
\label{eq_rep}
\widehat{\mathbf{E}}(S)= -i X(S), \qquad \widehat{\mathbf{H}}(S) = e^{ig\int_S \mathbf{B}^{(0)}}\;\;\;\;\mathbf{h}_{\partial S} 
\end{equation}
where the action of $X(S)$ was defined in \eqref{eq:III.3} and $h_{\partial S}$ is the usual holonomy. As the representation of $\mathbf{E}(S)$ is the usual one, it respects all the relations above that just contain $\mathbf{E}(S)$. Furthermore, the relations that just contain $\mathbf{H}_{S}$ are also satisfied, due to the fact that the additional phase in \eqref{eq_rep} is the exponential of a classical magnetic flux, and U(1) is Abelian. Finally, the commutation relations are represented by \eqref{eq_rep}, because the only change to the usual AL-representation is a phase.

The classical field $\mathbf{B}^{(0)}$ is, a priori, completely arbitrary. However, to keep the picture as closely analogous to that presented by the electric flux, we postulate that, similarly, $\mathbf{B}^{(0)}$ must consist of flux lines, carrying integer multiples of an elementary magnetic flux quantum. This postulate would need to be justified more thoroughly in the future. So, to specify $\mathbf{B}^{(0)}$ we need the data of a U(1) spin network $(\gamma_0, \underline{m})$. Then 
\begin{equation}
\mathbf{B}^{(0)a}(p)=g_m \sum_{e\in\gamma_0} m_e e^a(p)
\end{equation}
with
\begin{equation}
\label{eq_formfactor}
e^a(p)=\int \dot{e}^a(t) \delta^{(3)}(p,e(t)) \;\text{d}t 
\end{equation}
the form factor associated to the edge $e$. $g_m$ is a constant that sets the unit of magnetic charge. We will denote charge network states in this representation as 
\begin{equation}
\ket{\gamma, \underline{n}\,|\,\gamma_0, \underline{m}}, 
\end{equation}
where $(\gamma,\underline{n})$ is an arbitrary charge network, and $(\gamma_0, \underline{m})$ the one describing the background magnetic flux. In this representation, the magnetic fluxes through closed surfaces are diagonal, and given as the exponential of a flux operator $\widehat{P}_S$, and are trivial, 
\begin{equation}
\widehat{\mathbf{H}}(S)=\exp \left(ig \widehat{P}_S\right), \qquad \widehat{P}_S \ket{\gamma,\underline{n}\,|\,\gamma_0, \underline{m}}=
g_m\sum_{e\in\gamma_0} m_e\; I(S,e) \ket{\gamma,\underline{n}\,|\,\gamma_0, \underline{m}}=0 \qquad \text{ for } \partial S=0
\end{equation}
since the U(1) spin network $(\gamma_0, \underline{m})$ is gauge invariant. In other words, the construction so far adds non-trivial magnetic fluxes, but no magnetic charges yet. We have also not mentioned the fact that for the application to black holes, we will need $\Sigma$ to have an inner boundary. For LQG to couple to the CS theory on the boundary, charge network states have to be allowed to have open ends on the boundary $\partial \Sigma$. This leads to a nonzero electric flux through $\partial \Sigma$. In fact, because 
 $\widehat{Q}(S)=-\frac{1}{4\pi}\widehat{\mathbf{E}}(S)$ in the U(1) case,  any closed genus 0 surface containing the black hole will have the same flux. This means that the charge network state coupling to the CS theory states on the black hole horizon will have to be infinitely extended. The observables considered so far neither allow for open ends, nor for extension to infinity. There is a special kind of charge network that does both: The edges are paths starting on $\partial \Sigma$ and running out to infinity without ever branching. 
Such a string network is thus of the form 
\begin{equation}
f^{\text{(string)}}_{\sigma,\underline{n}^s}[\mathbf{A}]=\prod_{k=1}^{N} \mathbf{h}_{s_k}^{n_k}[\mathbf{A}], 
\end{equation}
where $\sigma=\{s_k\}$  is the set of strings running from $\partial\Sigma$ to infinity, in the sense that they have $I(s_k,S)=1$ where $S$ is any sphere containing $\partial \Sigma$. There are certainly issues with convergence with 
\begin{equation}
\mathbf{h}_{s_k} = e^{i \int_{s_k} \mathbf{A}}
\end{equation}
for classical connections not decaying sufficiently. We will not further investigate this here. We will try to avoid the corresponding issues in the quantum theory. However, this should be more carefully analyzed at some point.  

We will call states $f^{\text{string}}_{\sigma,\underline{n}^s}$ string network states, and the labels $(\sigma, \underline{n})$ a string network. We conjecture that any charge network state $F$ with open ends on $\partial\Sigma$ can be factorized as 
\begin{equation}
\label{eq_conjecture}
F= f^{\text{(string)}} \; f_0
\end{equation}
where $f$ is a gauge invariant charge network. If conjecture \eqref{eq_conjecture} holds, any reasonable charge net with open ends on $\partial\Sigma$ can be created by a product of holonomies along strings, and exponentiated magnetic fluxes $\mathbf{H}(S)$.
Thus, we will add to the quantum algebra operators 
\begin{equation}
\mathbf{h}_s \widehat{=} e^{i \int_s \mathbf{A}}
\end{equation}
where $s$ is a string running from $\partial\Sigma$ to infinity in the above sense.  We will not specify any further relations for the $\mathbf{h}_s$\footnote{This is mainly to avoid convergence problems. For example, two strings can intersect infinitely many times, and thus in principle form infinite charge networks with related questions about convergence and normalization.}, except for 
\begin{equation}
[\mathbf{E}(S), \mathbf{h}_{s}]= g I(S, s)\mathbf{h}_{s}. 
\end{equation}
A Hilbert space $\mathcal{H}_{\mathrm{YM}}$ analogous to the one of \eqref{eq_hilb} is defined by the inner product
\begin{equation}
\scpr{f^{\text{(string)}}_{\sigma, \underline{n}} \; f_0}{f'{}^{\text{(string)}}_{\sigma', \underline{n}'} \; f'_0}=
\scpr{f_0}{f'_0}_{\mathrm{YM,0}}\;\delta_{(\sigma, \underline{n}), (\sigma', \underline{n}')}. 
\end{equation}
This deals with the electric flux. For the magnetic flux, we have to extend the background magnetic field $\mathbf{B}^{(0)}$ by strings going from $\partial\Sigma$ to infinity in the same way as we have extended the holonomy observables $\mathbf{H}(S)$. To do that, we specify not only the data of a charge network $(\gamma_0,\underline{m})$, but also that of a string network
$(\sigma_0,\underline{m}^\sigma)$. Then 
\begin{equation}
\mathbf{B}^{(0)a}(p)=g_m\sum_{e\in\gamma} m_e e^a(p)+g_m\sum_{s\in \sigma} m^\sigma_s s^a(p). 
\end{equation}
with the form factors $e^a, s^a$ defined in \eqref{eq_formfactor}.

With these definitions, the electric and magnetic horizon charges become 
\begin{align}
\widehat{Q}_H \ket{\gamma,\underline{n},\sigma,\underline{n}^\sigma\,|\,\gamma_0,\underline{m}_0,\sigma_0,\underline{m}^\sigma_0}
&=g\left(\sum_{e\in \gamma} n_e I(e,H) +\sum_{s\in \sigma} n^\sigma_s\right)  \ket{\gamma,\underline{n},\sigma,\underline{n}^\sigma\,|\,\gamma_0,\underline{m}_0,\sigma_0,\underline{m}^\sigma_0},\\
\widehat{P}_H \ket{\gamma,\underline{n},\sigma,\underline{n}^\sigma\,|\,\gamma_0,\underline{m}_0,\sigma_0,\underline{m}^\sigma_0}
&=g_m\left(\sum_{e\in \gamma_0} m_e I(e,H) +\sum_{s\in \sigma_0} m^\sigma_{s}\right)  \ket{\gamma,\underline{n},\sigma,\underline{n}^\sigma\,|\,\gamma_0,\underline{m}_0,\sigma_0,\underline{m}^\sigma_0}.
\label{eq:YM_P}
\end{align}
We note again that the status of the electric and magnetic fluxes are somewhat different in the quantum theory: The electric ones can be changed by the operators $\widehat{\mathbf{H}}(S)$, $\mathbf{h}_s$, whereas the magnetic ones can not be changed, they label superselection sectors. Neverteless we will treat both in the same way when it comes to counting states to obtain entropy. 
\subsection{Kinematical Hilbert space}
With these preliminaries we are ready to quantize the phase space $(\Gamma,\Omega^{\Gamma})$ of the charged black hole. We define the kinematical Hilbert space of the system via the following tensor product
\begin{equation}
\mathcal{H}_{\mathrm{kin}}=\mathcal{H}_{\mathrm{grav}}\otimes\mathcal{H}_{\mathrm{YM}}\otimes\mathcal{H}^{\sigma_{+}}_{\mathrm{CS}}\otimes\mathcal{H}^{\sigma_{-}}_{\mathrm{CS}}
\label{eq:3.6.2.1}
\end{equation}
with $\mathcal{H}_{\mathrm{YM}}$ as defined previously and $\mathcal{H}_{\mathrm{grav}}=L^2(\bar{\mathcal{A}}_{\mathrm{SU}(2)},\mathrm{d}\mu_{\mathrm{AL}}^{\mathrm{SU}(2)})$ the Hilbert space of the quantized gravitational field in the bulk spanned by the spin-network states $\ket{\gamma,\underline{j},\underline{m}}$. In the general case, we assume that the magnetic charge of the black hole is vanishing, $P_H=0$. For U(1) we also sketch the possibility $P_H\neq 0$. 

The Gauss and diffeomorphism constraints on the gravitational and Yang-Mills phase space factorize according to the above decomposition and thus can be straightforwardly implemented on the respective Hilbert spaces leading to gauge-invariant spin- and charge-network states.
In contrast, the Hamiltonian constraint of the Yang-Mills field acts nontrivially on $\mathcal{H}_{\mathrm{grav}}$. It therefore yields a nontrivial coupling of the gravitational and YM degrees of freedom which, however, will not be discussed here. 

Finally, $\mathcal{H}^{\sigma_{\pm}}_{\mathrm{CS}}$ are the Hilbert spaces of the quantized $\mathrm{SU}(2)$ Chern-Simons gauge theories with Chern-Simons level \cite{Perez:2010pq}
\begin{equation}
k_{\pm}=\pm\frac{a_H}{4\pi\beta(\sigma_{-}^2-\sigma_{+}^2)}
\end{equation} 
As will be discussed below, the constraints (\ref{eq:3.4.15}), (\ref{eq:3.4.16}) as well as (\ref{eq:3.6.1.15}) are mapped to quantum boundary conditions in the quantum theory and lead to a coupling between the quantized bulk and horizon degrees of freedom. In particular, it follows that the Chern-Simons state space is given by $\mathrm{SU}(2)$ connections on $S^2$ which are flat except at the punctures $p\in\mathcal{P}$ where the spin-network graph of the quantized gravitational field intersects the black hole horizon. Let $\{j_p\}$ be the spin quantum numbers associated to edges intersecting the black hole at the punctures $p\in\mathcal{P}$. The quantum boundary conditions imply that at each puncture $p$, $j_p$ couples with two spin-quantum numbers $j_p^{\pm}$ assigned to the respective Chern-Simons Hilbert spaces $\mathcal{H}^{\sigma_{\pm}}_{\mathrm{CS}}$. Accordingly, $\mathcal{H}^{\sigma_{\pm}}_{\mathrm{CS}}$ are given by     
\begin{equation}
\mathcal{H}^{\sigma_{\pm}}_{\mathrm{CS}}\equiv\mathcal{H}^{\sigma_{\pm}}_{\mathrm{CS}}(\mathcal{P},\{j^{\pm}_p\})=\mathcal{H}^{\mathrm{SU}(2)}_{\mathcal{P},k_{\pm}}(\{j^{\pm}_p\})
\label{eq:3.6.2.2}
\end{equation}
with $\mathcal{H}^{\mathrm{SU}(2)}_{\mathcal{P},k_{\pm}}(\{j^{\pm}_p\})$ the Hilbert spaces of quantized $\mathrm{SU}(2)$ Chern-Simons theories with punctures $\mathcal{P}$ labeled by irreducibles $\{j_p^{\pm}\}$. These can be identified with the intertwiner subspaces of the tensor product representations $j_{p_1}^{\pm}\otimes\ldots\otimes j_{p_n}^{\pm}$ of a certain quantum deformation $\mathrm{SU}_q(2)$ of $\mathrm{SU}(2)$. Their dimensions are given by the celebrated Verlinde formula \cite{Engle:2011vf}
\begin{equation}
\mathrm{dim}\mathcal{H}^{\mathrm{SU}(2)}_{\mathcal{P},k_{\pm}}(\{j^{\pm}_p\})=\int_{0}^{2\pi}{\mathrm{d}\mu_k(\theta)\,\prod_{i=1}^n{\frac{\sin\left(d_i\frac{\theta}{2}\right)}{\sin\frac{\theta}{2}}}}:=\int_{0}^{2\pi}{\mathrm{d}\theta\,\frac{1}{\pi}\sin^2\left(\frac{\theta}{2}\right)\frac{\sin\left((2r+1)\frac{(k+2)\theta}{2}\right)}{\sin\left(\frac{(k+2)\theta}{2}\right)}\prod_{i=1}^n{\frac{\sin\left(d_i\frac{\theta}{2}\right)}{\sin\frac{\theta}{2}}}}
\label{eq:1.5.8}
\end{equation}
where $k:=k_{+}=-k_{-}$, $r:=\lfloor\frac{1}{k}\sum_{i=1}^n{j_i}\rfloor$ and $d_i:=\mathrm{dim}\,E^{j_i}=2j_i+1$. For $r=0$, this reduces the well-known integral formula for the dimension of the intertwiner subspace of the ordinary tensor product representation of standard $\mathrm{SU}(2)$. This is in particular the case for $k\rightarrow\infty$.
\subsection{Quantum boundary conditions}
As a next step, we need to implement the classical boundary conditions (\ref{eq:3.4.15}), (\ref{eq:3.4.16}) and (\ref{eq:3.6.1.15}) in the quantum theory. Concerning the pure gravitational degrees of freedom, this can be performed similarly as in \cite{Perez:2010pq}. Hence, we fix a certain puncture $p$ and a small disk $D_{\epsilon}(p)$ on the horizon of radius $\epsilon$ about $p$. Integrating both sides in (\ref{eq:3.6.1.15}) over $D_{\epsilon}(p)$ and taking the limit $\epsilon\rightarrow 0$, this yields\footnote{We assume that $\phi_1$, $\Psi_2$ and $c$ are continuous.}
\begin{equation}
\lim_{\epsilon\rightarrow 0}\int_{D_{\epsilon}(p)}{F(A_{\sigma_{\pm}})^i}=2\left(\Psi_2-2\|\phi_{1}\|^2+\frac{\pi}{a_H}\sigma_{\pm}^2+\frac{c}{2}\right)\lim_{\epsilon\rightarrow 0}E^i(D_{\epsilon}(p))
\label{eq:3.6.2.3}
\end{equation}    
where we used that $(\ast E)^i=\frac{1}{2}\tensor{\epsilon}{^i_{jk}}e^{j}\wedge e^{k}=\frac{1}{2}\Sigma^i$ at the horizon. If we then promote the classical expressions above into the corresponding operators in the quantum theory, we find at any puncture $p\in\mathcal{P}$
\begin{equation}
\widehat{J}^i_{\pm}(p):=\frac{k_{\pm}}{4\pi}\lim_{\epsilon\rightarrow 0}\int_{D_{\epsilon}(p)}{\widehat{F}(A_{\sigma_{\pm}})^i}=\frac{k_{\pm}}{4\pi}\left(\Psi_2-2\|\phi_{1}\|^2+\frac{\pi}{a_H}\sigma_{\pm}^2+\frac{c}{2}\right)\widehat{\Sigma}^i(p)
\label{eq:3.6.2.4}
\end{equation}
where the limit has to be understood to hold in the strong operator topology such that
\begin{equation}
\widehat{\Sigma}^i(p):=2\lim_{\epsilon\rightarrow 0}\widehat{E}^i(D_{\epsilon}(p))=\kappa\beta\widehat{J}^i(p)
\label{eq:3.6.2.5}
\end{equation}
It follows that $\widehat{J}_{\pm}$ indeed satisfy the commutation relations of angular momentum operators. The labels $j^{\pm}_p$ are associated to the eigenvalues of the respective Casimir operators $\widehat{J}^2_{\pm}(p)$. Hence, we arrive at the following quantum boundary conditions for the pure gravitational degrees of freedom
\begin{equation}
\widehat{J}^i_{\pm}(p)=\pm\frac{\frac{a_H}{\pi}\left(\Psi_2-2\|\phi_1\|^2+\frac{c}{2}\right)+\sigma_{\pm}^2}{\sigma_{-}^2-\sigma_{+}^2}\widehat{J}^i(p)
\label{eq:3.6.2.6}
\end{equation}
which have to hold at any puncture $p\in\mathcal{P}$. Let us quantize the constraints (\ref{eq:3.4.15}) and (\ref{eq:3.4.16}) for the Yang-Mills fields in a similar way. Therefore, we fix a point $p\in\mathcal{P}$, choose a disk $D_{\epsilon}(p)$ of radius $\epsilon$, integrate (\ref{eq:3.4.15}) and (\ref{eq:3.4.16}) over $D_{\epsilon}(p)$ and take the limit $\epsilon\rightarrow 0$. In doing so, we find 
\begin{align}
\lim_{\epsilon\rightarrow 0}Q(D_{\epsilon}(p))&=\lim_{\epsilon\rightarrow 0}-\frac{1}{4\pi}\int_{D_{\epsilon}(p)}{\|\mathbf{E}\|}\nonumber\\
&=\frac{1}{2\pi}\|\mathrm{Re}(\phi_1)\|\lim_{\epsilon\rightarrow 0}\int_{D_{\epsilon}(p)}{\mathrm{vol}_{S^2}}=\frac{1}{2\pi}\|\mathrm{Re}(\phi_1)\|\lim_{\epsilon\rightarrow 0}\mathrm{vol}(D_{\epsilon}(p))
\label{eq:3.6.2.7}
\end{align}
and analogously for (\ref{eq:3.4.15}). Hence, by again promoting the classical expressions into operators, this yields the following quantum boundary conditions for the quantized Yang-Mills fields 
\begin{equation}
\widehat{Q}(p):=\lim_{\epsilon\rightarrow 0}\widehat{Q}(D_{\epsilon}(p))=\frac{1}{2\pi}\|\mathrm{Re}(\phi_1)\|\widehat{a}_{H}(p)
\label{eq:3.6.2.8}
\end{equation}
and for the case of U(1) also
\begin{equation}
\widehat{P}(p):=\lim_{\epsilon\rightarrow 0}\widehat{P}(D_{\epsilon}(p))=\frac{1}{2\pi} \mathrm{Im}(\phi_1)\; \widehat{a}_{H}(p)
\label{eq:3.6.2.9}
\end{equation}
where the limit again holds in the strong sense. The operator $\widehat{a}_{H}(p)$ is the ordinary area operator in quantum geometry associated to the puncture $p$ acting via
\begin{equation}
\widehat{a}_{H}(p)\ket{\vec{j},\vec{m}}_{\mathcal{P}}=\frac{\kappa}{2}\beta\sqrt{j_p(j_p+1)}\ket{\vec{j},\vec{m}}_{\mathcal{P}}
\label{eq:3.6.2.10}
\end{equation}
The constraints (\ref{eq:3.6.2.6}), (\ref{eq:3.6.2.8}) as well as (\ref{eq:3.6.2.9}) are the quantum boundary conditions that have to be imposed on the kinematical Hilbert space $\mathcal{H}_{\mathrm{kin}}$ of the quantized charged black hole.\\
In order to implement the boundary conditions (\ref{eq:3.6.2.6}) of the pure gravitational degrees of freedom, let us re-express them in terms of the constraints $\widehat{C}^i(p)$ and $\widehat{D}^i(p)$ defined as
\begin{equation}
\widehat{D}^i(p)=\widehat{J}^i(p)+\widehat{J}_{+}^i(p)+\widehat{J}_{-}^i(p)=0
\label{eq:3.6.3.1}
\end{equation}
and
\begin{equation}
\widehat{C}^i(p)=\widehat{J}_{+}^i(p)-\widehat{J}_{-}^i(p)-\alpha\widehat{J}^i(p)=0
\label{eq:3.6.3.2}
\end{equation}
at any puncture $p\in\mathcal{P}$ where $\alpha$ now takes the form 
\begin{equation}
\alpha=\frac{\frac{a_H}{\pi}d+(\sigma_{-}^2+\sigma_{+}^2)}{\sigma_{-}^2-\sigma_{+}^2}
\label{eq:3.6.3.3}
\end{equation}
with the new distortion parameter $d:=2(\Psi_2-2\|\phi_1\|^2)+c$. As can be seen, these constraints just have the same form as in \cite{Perez:2010pq,DiazPolo:2011np} only with slightly modified parameters $\alpha$ and $d$. Thus, the implementation of these constraints can be performed in complete analogy to \cite{Perez:2010pq}. Let us only state the result: Since the algebra generated by $\widehat{D}^i(p)$ closes under the commutator, these constraints are first-class and therefore can be implemented strongly. The $\widehat{D}^i(p)$'s, however, do not close. Thus, one instead requires that the corresponding master constraint $\widehat{C}^2(p)$ only holds in a weak sense, that is, in the large $j$ limit. As it turns out, this is equivalent to requiring  
\begin{equation}
j_p=j_p^{+}+j_p^{-}
\label{eq:3.6.3.12}
\end{equation}  
at every puncture $p\in\mathcal{P}$. Furthermore, it follows that $\alpha$ becomes a local operator in the quantum theory given by  
\begin{equation}
\widehat{\alpha}(p)=\frac{\widehat{J}_{-}^2(p)-\widehat{J}_{+}^2(p)}{\widehat{J}^2(p)}
\label{eq:3.6.3.13}
\end{equation}
In particular, according to (\ref{eq:3.6.3.3}), the classical quantity $d$ and thus $\Psi_2$, $\phi_1$ and $c$ all have to be interpreted as local operators $\widehat{d}(p)$, $\widehat{\Psi}_2(p)$, $\widehat{\phi}_1(p)$ and $\widehat{c}(p)$ associated to punctures $p\in\mathcal{P}$ such that
\begin{align}
\widehat{d}(p)&=2(\widehat{\Psi}_2(p)-2\|\widehat{\phi}_1(p)\|^2)+\widehat{c}(p)\nonumber\\
&=\frac{\pi}{a_H}\left(\frac{(1-\sigma_{+}^2)(\widehat{J}_{-}(p)^2-\widehat{J}_{+}^2(p))-(1+\sigma_{+}^2)\widehat{J}^2(p)}{\widehat{J}^2(p)}\right)
\label{eq:3.6.3.14}
\end{align}
where the free parameter $\sigma_{-}$ was set to $\sigma_{-}=1$ in order to include the spherical symmetric limit. The operator $\widehat{d}$ encodes the degree of distortion of the black hole in the quantum theory and commutes with all operators acting on the horizon. In analogy to the classical theory, spherical symmetry of the system is encoded in terms of eigenstates of the distortion operator to the eigenvalue $-2\pi/a_H$. These are characterized by $j_p^{+}=j_p$ and $j_p^{-}=0$ called spherically symmetric states.\\  
So far, we only considered the quantum boundary conditions of the pure gravitational system. But, to construct the full physical Hilbert space, we still need to implement the constraints (\ref{eq:3.6.2.8}) and (\ref{eq:3.6.2.9}) for the YM degrees of freedom. Therefore, as recently observed, we note that the classical quantity $\phi_1$ is associated to a local operator in the quantum theory. Hence, since $\widehat{a}_{H}(p)\neq 0$ at the punctures, the boundary conditions (\ref{eq:3.6.2.8}) and (\ref{eq:3.6.2.9}) have to be interpreted as defining equations for the operator $\widehat{\phi}_1$ which satisfies
\begin{equation}
\widehat{\phi}_e(p):=\|\mathrm{Re}(\widehat{\phi}_1)(p)\|=2\pi\frac{\widehat{Q}(p)}{\widehat{a}_{H}(p)}
\label{eq:3.6.3.16}
\end{equation}
where the norms are dropped in the special case $G=\mathrm{U}(1)$. In addition, in this case we also have 
\begin{equation}
\widehat{\phi}_m(p):=\mathrm{Im}(\widehat{\phi}_1)(p)=2\pi\frac{\widehat{P}(p)}{\widehat{a}_{H}(p)}
\label{eq:3.6.3.16b}
\end{equation}
According to \eqref{eq:3.6.3.16}, \eqref{eq:3.6.3.16b}, we can interpret $\widehat{\phi}_e$ and $\widehat{\phi}_m$ as local charge-density operators. In case of spherical symmetry, we saw by (\ref{eq:II.32}) and (\ref{eq:II.34}) that, classically, $\|\phi_1\|$ is completely determined by macroscopic quantities, i.e. charge and area of the black hole. Hence, in the quantum theory we have to require that spherically symmetric states are eigenstates of the operators $\widehat{\phi}_e(p)$ and $\widehat{\phi}_m(p)$ with eigenvalue $\frac{2\pi}{a_H}Q_H$ and $\frac{2\pi}{a_H}P_H$, respectively, fixing the quantum numbers of the charge networks at the horizon. For distorted black holes, however, there is no classical restriction for $\phi_1$. Hence, in this case, equation (\ref{eq:3.6.3.16}) just yields two boundary operators determining the local charge-density at the horizon which can take any values and thus, in particular, leads to no further restriction of physical states.

\section{Entropy} 
\label{se_entropy}
\subsection{Distorted horizons}
After we have constructed the quantum theory for generic non-rotating charged black holes, we now want to go over to the computation of the black hole entropy. Here and in the following, we therefore restrict for non-Abelian gauge groups to the situation that the black hole horizon does not carry a nonzero total magnetic charge $P_H$. However, let us remark that spherically symmetric black holes with purely electric charge may not be stable for general Einstein-Yang-Mills theories \cite{Corichi:2000dm}. Hence, in section \ref{ap_mag}, we sketched a possibility how magnetic charges can be included into the theory if $G=\mathrm{U}(1)$. However, at this stage, it is not clear yet how this construction can be generalized to arbitrary non-Abelian gauge theories. More details on this will be reported elsewhere \cite{ESS}. 

For the computation of the entropy we work with the microcanonical ensemble and, as usual, first consider the entropy to be associated to the pure horizon degrees of freedom of the quantized system. More precisely, given the total area $a_H$ and charge $Q_H$ of the black hole, we count for an arbitrary set of punctures $\mathcal{P}$, arising from the intersection of the product spin-network states $f_{\gamma,\vec{j},\vec{m}}\otimes f_{\gamma,\vec{\lambda}}$ with the horizon, the number of surface states $\mathcal{N}(a_H,Q_H)$, that is, states in $\mathcal{H}_{\mathrm{CS}}^{\sigma_{+}}(\mathcal{P},\{j_p^{+}\})\otimes\mathcal{H}_{\mathrm{CS}}^{\sigma_{-}}(\mathcal{P},\{j_p^{-}\})$ which satisfy the constraints (\ref{eq:3.6.3.12}) as well as 
\begin{equation}
a_H=4\pi\beta\sum_{p\in\mathcal{P}}{\sqrt{(d_p^{+}+d_p^{-})^2-1}}
\label{eq:3.6.4.5}
\end{equation}     
and 
\begin{equation}
Q_H=\sum_{p\in\mathcal{P}}{Q(\lambda_p)}
\label{eq:3.6.4.6}
\end{equation}
where we used that $4j(j+1)=(d-1)(d+1)=d^2-1$ with $d:=2j+1$. The entanglement entropy $S_{\mathrm{BH}}$ of the charged black hole is then defined as the corresponding von Neumann entropy of a maximally mixed state of the quantum system such that
\begin{equation}
S_{\mathrm{BH}}=\ln\mathcal{N}(a_H,Q_H)
\label{eq:3.5.2.11}
\end{equation}
For purely distorted configurations, i.e. $\widehat{d}(p)\neq -2\pi/a_H$, we have seen above that, since, classically, $\phi_1$ is not restricted to any specifc values in this case, the charge-density operator $\widehat{\phi}_1(p)$ does not lead to any further constraint for the
physical states. But then condition (\ref{eq:3.6.4.6}) can be solved trivially and, in particular, completely independent of the choice of the purely gravitational degrees of freedom. Consequently, the number of surface states solely depends on the total area but not on the charge of the black hole so that the corresponding entropy coincides with the uncharged case. Hence, nothing changes for purely distorted black holes as compared to the standard calculation without matter \cite{Perez:2010pq}. This is however no longer true in the spherical symmetric limit, to which we turn next. 
\subsection{The spherical symmetric limit}
Since spherically symmetric states are eigenstates of the charge-density operator $\widehat{\phi}_e$ with eigenvalue $2\pi Q_H/a_H$, it follows from boundary condition (\ref{eq:3.6.3.16}) that the surface states need to obey the following additional constraints\footnote{Note that $j_p^{+}=j$ and $j_p^{-}=0$ for spherically symmetric states.}  
\begin{equation}
2\pi\frac{Q_H}{a_H}=2\pi\frac{Q(\lambda_p)}{4\pi\beta\sqrt{j_p(j_p+1)}}\Leftrightarrow Q(\lambda_p)=\frac{Q_H}{a_H}4\pi\beta\sqrt{j_p(j_p+1)}
\label{eq:IV.4}
\end{equation}
If $Q_H=0$ this condition just means that $Q(\lambda_p)=0$ at every puncture $p\in\mathcal{P}$. Thus, the uncharged limit is included in the theory. However, for $Q_H\neq 0$ this condition has dramatic consequences. In fact, for physical black holes the total area is much larger than the charge so that the factor $Q_H/a_H$ in (\ref{eq:IV.4}) becomes very small. But, the spectrum of the charge operator $\widehat{Q}$ of the quantized Yang-Mills theory is discrete and bounded from below. Thus, there will be too few surface states, if any, which satisfy these boundary conditions. This is a particularity of the spherical symmetric limit as it requires a direct coupling between the spectra of the area and charge operators which can have completely different properties. We will however see in section \ref{AnCon} that this problem in the spherically symmetric limit can be resolved if one allows for an imaginary $\beta$.\\ 
Another possibility to allow for a non-pathological limit with spherical symmetry exists for the case of $G=U(1)$. In that case, one can obtain a quantization in which the charge operator 
\eqref{eq_QU1} obtains a (discrete) spectrum which is equal to $\mathbb{R}$. To that end, one considers states 
\begin{equation}
\braket{\mathbf{A}|\gamma,\underline{r}}=h_{e_1}(\mathbf{A})^{r_{e_1}}\dots h_{e_k}(\mathbf{A})^{r_{e_k}}\qquad \text{for}\quad\gamma=\{e_1,\ldots,e_k\} \quad \underline{r}=\{ r_{e_1}, r_{e_2}, \ldots\} \in \mathbb{R}^{|\gamma|}
\label{eq_RBohr}
\end{equation}
with the inner product given by
\begin{equation}
\braket{\gamma,\underline{r}|\gamma',\underline{r}'}=\delta_{\gamma,\gamma'}\delta_{\underline{r},\underline{r}'}.  
\end{equation}
The charge operator is then given by 
\begin{equation}
\widehat{Q}(S)\ket{\gamma,\underline{r}}=g\left(\sum_{e\cap S}{r_e}\right)\ket{\gamma,\underline{r}}
\end{equation} 
In this case, charge quantization is not automatic anymore, but has to be imposed at each vertex, by requiring 
\begin{equation}
\sum_{e\text{ at }v} r_e=n_v \in \mathbb{Z}
\end{equation}
at each vertex $v$ of $\gamma$. For this quantization, 
$\mathcal{H}_{\mathrm{EM}}=L^2(\bar{\mathcal{A}}_{\overline{\mathbb{R}}_\text{B}},\mathrm{d}\mu_{\mathrm{AL}}^{\overline{\mathbb{R}}_\text{B}})$, with $\overline{\mathbb{R}}_\text{B}$ the Bohr-compactification of $\mathbb{R}$.\footnote{For details on $\overline{\mathbb{R}}_\text{B}$ see for example \cite{Ashtekar:2003hd,Fewster:2008sr}.} 

Since electric flux can take any real value in this quantization, the problems with \eqref{eq:IV.4} are resolved and one obtains the usual result for the entropy. For details see \cite{KE_master}. Since this quantization is not particularly natural for a U(1) gauge theory, and because an analogous option does not seem to be available in general for other structure groups, we will not pursue this option further in this work. 

\subsection{Inclusion of YM DOF}
So far, we considered the entropy to be given by the number of surface states which can exist under specification of the total area and charge of the black hole. But, alternatively, one could also think of the Yang-Mills degrees of freedom at the horizon to additionally contribute to the total entropy. In case of spherical symmetry, condition (\ref{eq:II.32}) and (\ref{eq:II.34}) for the charge-density imply that, according to (\ref{eq:IV.4}), the charge-network labels are completely fixed by the spin quantum numbers. Due to this direct coupling, the inclusion of the Yang-Mills degrees of freedom does not increase the number of states in the state counting and everything remains unchanged. This argument, however, does not apply to general distorted black holes. In fact, since the charge-density operator is not restricted to any specific values in this case, (\ref{eq:3.6.4.6}) remains the only condition that needs to be imposed on the charge-network states.\\
In the following, we want to concentrate on the special case $G=\mathrm{U}(1)$ and first assume that $P_H=0$. We will then also extend the analysis to $P_H\neq 0$ and finally briefly comment on the non-Abelian case at the end of this section. For $G=\mathrm{U}(1)$ the charge quantum numbers $n$ can be both positive and negative. Hence, there exist infinitely many states that satisfy condition (\ref{eq:3.6.4.6}) and, consequently, the number of states diverges. However, by specifying a certain type of regularization, it can be shown that, in a certain sense, this number is still proportional to the black hole area.\\
Let us therefore introduce a regularization by fixing a natural number $N_{\max}\in\mathbb{N}$ and only counting charge quantum numbers $n$ for which $|n|\leq N_{\max}$. Given the total area $a_H$ and charge $Q_H=Ng$ of the black hole, we thus need to count, for a fixed number $P:=|\mathcal{P}|\in\mathbb{N}_0$ of punctures $p\in\mathcal{P}$, the number $A_P(N)$ of integers $n_p\in\mathbb{Z}$ such that $|n_p|\leq N_{\max}$ and 
\begin{equation}
\sum_{p\in\mathcal{P}}{n_p}=N
\label{eq:3.6.5.1}
\end{equation}
Let us determine $A_P(N)$ by induction on $P$. Obviously, for $P=0$ one has $A_0(N)=0$. Furthermore, for $P=1$ there is only one possible state so that $A_1(N)=1$. If $P=2$, it follows from (\ref{eq:3.6.5.1}) that the charge quantum number of one particular puncture already fixes the charge quantum number of the remaining one. Hence, $A_2(N)$ coincides with the number of integers $k\in\mathbb{Z}$ such that $|k|\leq N_{\max}$ as well as $|N-k|\leq N_{\max}$. It follows that $A_2(N)$ is given by    
\begin{equation}
A_2(N)=\max\{2N_{\max}+1-|N|,0\}
\label{eq:3.6.5.2}
\end{equation} 
where the $\max$-function is needed, since $A_2(N)=0$ if $2N_{\max}<|N|$. For $P>2$, due to (\ref{eq:3.6.5.1}), the number $A_P(N)$ can be determined via the $A_{P-1}(N')$ for certain integers $N'\in\mathbb{Z}$ using the recurrence relation
\begin{equation}
A_P(N)=\sum_{k=-N_{\max}}^{N_{\max}}{A_{P-1}(N-k)}
\label{eq:3.6.5.3}
\end{equation}  
Unfortunately, this recurrence relation cannot be solved analytically. However, since the regulator $N_{\max}$ is sent to infinity at the end of the regularization procedure, let us estimate $A_P(N)$ in the limit $N_{\max}\gg N$. In this case, it follows from (\ref{eq:3.6.5.2}) that $A_2(N)\approx 2N_{\max}$. Hence, in the limit $N_{\max}\rightarrow\infty$, the recurrence relation (\ref{eq:3.6.5.3}) is approximately solved for 
\begin{equation}
A_{P}(N)\approx(2N_{\max})^{P-1}
\label{eq:3.6.5.4}
\end{equation}
where $P\geq 1$. With the aid of this formula, let us go over to the computation of the black hole entropy. Using (\ref{eq:1.5.8}) for $N_k(\{j^{\pm}_p\}):=\mathrm{dim}\,\mathcal{H}_{\mathcal{P},k}^{\mathrm{SU}(2)}(\{j_p^{\pm}\})$, the dimension of the Hilbert space of the quantized $\mathrm{SU(2)}$ Chern-Simons theory, we find, together with the boundary condition (\ref{eq:3.6.4.5}) and constraint (\ref{eq:3.6.3.12}), that the inclusion of electromagnetic degrees of freedom leads to the following formula for the total number $\mathcal{N}(a)$ of black hole states, setting $a=\frac{a_H}{4\pi\beta}$,\footnote{Note that the spherically symmetric limit is not excluded in the state counting, although we have noted before that there are some inconsistencies in this limit. But, perhaps another way to deal with this limit could be found (see fo example section \ref{AnCon} below). Furthermore, the inclusion of the spherically symmetric states just slightly changes the further considerations. Thus we will include the states here.} 
\begin{equation}
\mathcal{N}(a)=\sum_{n=0}^{\infty}{\sum_{d^{\pm}_{p_1},\dots,d^{\pm}_{p_n}=1}^{k+1}{\delta\left(a-\sum_{i=1}^n{\sqrt{(d_{p_i}^{+}+d_{p_i}^{-})^2-1}}\right)N_k(\{j_p^{+}\})N_k(\{j_p^{-}\})}}(2N_{\max})^{n-1}
\label{eq:3.6.5.5}
\end{equation}  
where the first sum runs over the total number $n$ of punctures labeled by irreducible representations $j_p^{\pm}$ of $\mathrm{SU}(2)$ of dimension $d_p^{\pm}=2j_p^{\pm}+1$. This number grows exponentially like $\mathcal{N}(a)\sim\exp(s_0a)$. As usual, in order to determine the value of $s_0$, let us apply the Laplace transform on (\ref{eq:3.6.5.5}). In a similar way as in \cite{Engle:2011vf}, we then find
\begin{align}
\tilde{\mathcal{N}}(s)=&\int_0^{\infty}{\mathrm{d}a\,\mathcal{N}(a)e^{-sa}}\nonumber\\
=&\sum_{n=0}^{\infty}{\sum_{\{d^{\pm}_p\}}^{k+1}{\int_0^{2\pi}{\mathrm{d}\theta^{+}\,\mu_k(\theta^{+})\int_0^{2\pi}{\mathrm{d}\theta^{-}\,\mu_k(\theta^{-})}}}}\left(\prod_{i=1}^{n}{\frac{\sin\left(\frac{d^{+}_{p_i}}{2}\theta^{+}\right)}{\sin\frac{\theta^{+}}{2}}\frac{\sin\left(\frac{d^{-}_{p_i}}{2}\theta^{-}\right)}{\sin\frac{\theta^{-}}{2}}e^{-s\sqrt{(d_{p_i}^{+}+d_{p_i}^{-})^2-1}}}\right)(2N_{\max})^{n-1}
\label{eq:3.6.5.6}
\end{align}
Interchanging the integrals with the sum over $d_p^{\pm}$ and subsequently rearranging the sum and product, this yields   
\begin{align}
\tilde{\mathcal{N}}(s)=&\sum_{n=0}^{\infty}{\int_0^{2\pi}{\mathrm{d}\theta^{+}\,\mu_k(\theta^{+})\int_0^{2\pi}{\mathrm{d}\theta^{-}\,\mu_k(\theta^{-})}}}\left(\sum_{d^{\pm}=1}^{k+1}{\frac{\sin\left(\frac{d^{+}}{2}\theta^{+}\right)}{\sin\frac{\theta^{+}}{2}}\frac{\sin\left(\frac{d^{-}}{2}\theta^{-}\right)}{\sin\frac{\theta^{-}}{2}}e^{-s\sqrt{(d^{+}+d^{-})^2-1}}}
\right)^n(2N_{\max})^{n-1}\nonumber\\
=&\frac{1}{2N_{\max}}\sum_{n=0}^{\infty}{\int_0^{2\pi}{\mathrm{d}\theta^{+}\,\mu_k(\theta^{+})\int_0^{2\pi}{\mathrm{d}\theta^{-}\,\mu_k(\theta^{-})}}}\left(\sum_{d^{\pm}=1}^{k+1}{\frac{\sin\left(\frac{d^{+}}{2}\theta^{+}\right)}{\sin\frac{\theta^{+}}{2}}\frac{\sin\left(\frac{d^{-}}{2}\theta^{-}\right)}{\sin\frac{\theta^{-}}{2}}e^{\ln(2N_{\max})-s\sqrt{(d^{+}+d^{-})^2-1}}   }
\right)^n\nonumber\\
=&\frac{1}{2N_{\max}}\int_0^{2\pi}{\mathrm{d}\theta^{+}\,\mu_k(\theta^{+})\int_0^{2\pi}{\mathrm{d}\theta^{-}\,\mu_k(\theta^{-})\tilde{D}(\theta^{+},\theta^{-},s)^{-1}}}
\label{eq:3.6.5.7}
\end{align}
where the well-known limit of the geometric series was used, yielding the function $\tilde{D}(\theta^{+},\theta^{-},s)$ given by
\begin{equation}
\tilde{D}(\theta^{+},\theta^{-},s)=1-\sum_{d^{\pm}=1}^{k+1}{\frac{\sin\left(\frac{d^{+}}{2}\theta^{+}\right)}{\sin\frac{\theta^{+}}{2}}\frac{\sin\left(\frac{d^{-}}{2}\theta^{-}\right)}{\sin\frac{\theta^{-}}{2}}e^{\ln(2N_{\max})-s\sqrt{(d^{+}+d^{-})^2-1}}}
\end{equation} 
The critical exponent $s_0=:\pi\beta_M^{k,N_{\max}}$ corresponds to the highest root of this function. This is the case for $\theta^{+}=\theta^{-}=0$ and thus is the unique solution of 
\begin{equation}
\frac{1}{2N_{\max}}=\sum_{d^{\pm}=0}^k{(d^{+}+1)(d^{-}+1)e^{-\pi\beta_M^{k,N_{\max}}\sqrt{(d^{+}+d^{-}+1)(d^{+}+d^{-}+3)}}}
\label{eq:3.6.5.8}
\end{equation} 
The number of black hole states $\mathcal{N}(a)$ then grows like $\mathcal{N}(a)=(2N_{\max})^{-1}\exp(\pi\beta_M^{k,N_{\max}}a)$ such that the corresponding entropy becomes
\begin{equation}
S_{\mathrm{BH}}(a_H)=\frac{\beta_M^{k,N_{\max}}}{\beta}\frac{a_H}{4l_p^2}-\ln(2N_{\max})
\label{eq:3.6.5.9}
\end{equation}
To find an approximate expression for $\beta_M^{k,N_{\max}}$, note that, according to (\ref{eq:3.6.5.8}), $\beta_M^{k,N_{\max}}$ tends to infinity in the limit $N_{\max}\rightarrow\infty$. Thus, in this limit, the sum in (\ref{eq:3.6.5.8}) becomes dominated by the $d_{\pm}=0$ term such that we can approximately write
\begin{align}
\frac{1}{2N_{\max}}\approx e^{-\pi\beta_M^{k,N_{\max}}\sqrt{3}}\Rightarrow\ln(2N_{\max})=\pi\beta_M^{k,N_{\max}}\sqrt{3}
\label{eq:3.6.5.10}
\end{align}
This is solved for
\begin{equation}
\beta_M^{k,N_{\max}}=\frac{\ln(2N_{\max})}{\pi\sqrt{3}}=\frac{\ln 2}{\pi\sqrt{3}}+\frac{\ln N_{\max}}{\pi\sqrt{3}}
\label{eq:3.6.5.11}
\end{equation}
which interestingly holds for any value $k$ of the Chern-Simons theory. Thus, if one includes the electromagnetic degrees of freedom in the state counting and performs such type of regularization, we can conclude that the black hole entropy is still proportional to the black hole area in leading order with proportionality factor explicitly depending on the regulator.

This still remains valid for $P_H\neq 0$. In fact, in this case, since magnetic charges can also be positive and negative, one again needs to introduce a regulator $M_{\mathrm{max}}$ for the counting of magnetic quantum numbers which leads to an additional factor $(2M_{\mathrm{max}})^{n-1}$ in formula (\ref{eq:3.6.5.5}) for the total number $\mathcal{N}(a_H,Q_H,P_H)$ of surface matter and gravity degrees of freedom. Hence, the computation is completely analogous to the case $P_H=0$ just with $2N_{\mathrm{max}}$ replaced by $4N_{\mathrm{max}}M_{\mathrm{max}}$
such that one ends up with
\begin{equation}
S_{\mathrm{BH}}(a_H)=\frac{\beta_M^{k,N_{\max},M_{\max}}}{\beta}\frac{a_H}{4l_p^2}-\ln(2N_{\max})-\ln(2M_{\max})
\label{eq:IV.17}
\end{equation} 
where
\begin{equation}
\beta_M^{k,N_{\max},M_{\max}}=\frac{\ln(4N_{\max}M_{\max})}{\pi\sqrt{3}}.
\end{equation}
Finally, let us briefly comment on the general situation of a non-Abelian gauge group. Here, the charge operator $\widehat{Q}$ is by definition positive definite. In contrast to $\mathrm{U}(1)$, the number of possible matter degrees of freedom is thus bounded by the total charge $Q_H$ of the black hole and therefore a regulator is not needed. We assume that the number of YM DOF again grows exponentially, at least in the large $Q_H$ limit. As for $\mathrm{U}(1)$ we would then end up with a similar result as in (\ref{eq:3.6.5.10}). 

Overall, the result that we find when including the quantum states corresponding to the distribution of electric and magnetic fluxes is very similar to what has been found for entanglement entropy across boundaries in matter ground states \cite{Bombelli:1986rw,Srednicki:1993im}. The entanglement entropy turns out to be proportional to area, but with a divergent, and hence regulator dependent prefactor. Thus we can perhaps read equation \eqref{eq:3.6.5.11} as giving a decomposition of total entropy \eqref{eq:3.6.5.9} in a finite part coming from the geometry, and an divergent one from the YM matter. 

\subsection{Analytic continuation}
\label{AnCon}
\subsubsection{General considerations}
As we have seen at the beginning of this chapter, the spherical symmetric limit requires a direct coupling between the quantized Yang-Mills and gravitational degrees of freedom. Due to the different charge and area spectra, this leads to a strong restriction of admissible surface states so that the Bekenstein-Hawking law is not recovered. One possible resolution of this problem would be if any of the operators $\widehat{Q}$ and $\widehat{a}_H$ had a continuous spectrum.\\
As already stated earlier, in the case $G=\mathrm{U}(1)$, this can be achieved for instance by quantizing the system by means of the Bohr compactification of the real line $\mathbb{R}_{\mathrm{Bohr}}$ as well known from loop quantum cosmology. The charge-network edges are then labeled by arbitrary real numbers and the boundary condition (\ref{eq:3.6.5.1}) is trivially solved. Consequently, the number of surface states remains unchanged the Bekenstein-Hawking law is recovered.\\
Another possibility, which holds for any YM theory, is to think of $\beta$ as a regulator of the theory and to set $\beta=\pm i$ at the end of the calculations, i.e. to perform some kind of Wick rotation. This was also discussed in \cite{BenAchour:2016mnn} in the context of rotating black holes. Here, we want to apply the techniques developed in \cite{Achour:2014eqa} to compute the entropy of a spherically symmetric charged black hole. We therefore again assume that the black hole solely carries electric charge and set $P_H=0$. The great advantage of this approach is that in this case the area spectrum indeed becomes continuous parametrized by some real number $s$ labeling the edges of the spin-network such that
\begin{equation}
\widehat{a}_H(p)=4\pi\sqrt{1+s_p^2}
\end{equation}
at any puncture $p\in\mathcal{P}$. Since spherically symmetric states are eigenstates of $\widehat{\phi}_e(p)$ to eigenvalues $\frac{2\pi}{a_H}Q_H$, it then follows
\begin{equation}
2\pi\frac{Q_H}{a_H}=2\pi\frac{Q(\lambda_p)}{4\pi\sqrt{1+s_p^2}}\Leftrightarrow Q(\lambda_p)=\frac{Q_H}{a_H}4\pi\sqrt{1+s_p^2}
\label{eq:IV.18}
\end{equation}
where $Q(\lambda_p)$ are the eigenvalues of the charge operator depending on the highest weights $\lambda_p$ of the compact gauge group $G$, associated to the charge-network edges. This yields a direct coupling between the charge-network labels $\lambda_p$ and the quantum numbers $s_p$ according to
\begin{equation}
s_p=\sqrt{\left(Q(\lambda_p)\frac{a_H}{4\pi Q_H}\right)^2-1}
\end{equation} 
where, of course, only $\lambda$'s are allowed for which the above expression makes sense. For physical black holes, the total area is much larger than the charge. In this case, we can approximately write
\begin{equation}
s_p\approx Q(\lambda_p)\frac{a_H}{4\pi Q_H}
\label{eq:Q1}
\end{equation} 
For imaginary $\beta$ the level $k$ of the Chern-Simons theory as well as the spin quantum numbers also become complex numbers. Formula (\ref{eq:1.5.8}) for the dimension of the quantized $\mathrm{SU}(2)$ Chern-Simons theory then becomes ill-defined. To cure this, one performs an analytic continuation of the dimension formula corresponding to some kind of Wick rotation. In this way, in \cite{Achour:2014eqa} one has found following formula in the large $k$ limit 
\begin{equation}
I_{\infty}(\{s_p\})=\frac{1}{i\pi}\oint_C{\mathrm{d}z\,\sinh^2(z)\prod_{p=1}^n{\frac{\sinh(is_p z)}{\sinh(z)}}}
\label{eq:IV.21}
\end{equation}
where $C$ is some contour in the complex plane encircling the poles of the meromorphic function in the integrand. With this formula, let us compute the entropy of the charged black hole. We again assume that the entropy of purely geometric origin and just depends on the surface degrees of freedom. For a given charge $Q_H$ and area $a_H$ of the black hole, we thus need to count for an arbitrary number $n$ of punctures $\mathcal{P}$ the number of surface states such that there are spin- and charge-network states labeled by $s_p$ and $\lambda_p$ with
\begin{equation}
a_H=4\pi\sum_{p=1}^n{\sqrt{1+s_p^2}}\quad\mathrm{and}\quad Q_H=\sum_{p=1}^n{Q(\lambda_p)}
\label{eq:IV.22}
\end{equation}
According to (\ref{eq:Q1}), the spin-network labels $s_p$ are completely fixed by the charge eigenvalues $Q(\lambda_p)$. In contrast to $s_p$, these eigenvalues can only take discrete values. Thus, counting surface states is equivalent to counting the $\lambda_p$'s  such that the second equation in (\ref{eq:IV.22}) holds. By (\ref{eq:Q1}) the remaining condition in (\ref{eq:IV.22}) is then automatically satisfied. Hence, together (\ref{eq:IV.21}) we find 
\begin{equation}
\mathcal{N}(a_H,Q_H)=\sum_{n=1}^{\infty}{\sum_{\{\lambda_p\}}{\delta\left(Q_H-\sum_{p=1}^{n}{Q(\lambda_p)}\right)I_{\infty}(\{s_p\})}}
\label{eq:Q2}
\end{equation}
By (\ref{eq:Q1}), we expect that for large areas the spin-network quantum numbers $s_p$ also become very large. In this limit, it has been shown in \cite{Achour:2014eqa} that $I_{\infty}(\{s_p\})$ approximately takes the form
\begin{equation}
I_{\infty}(\{s_p\})\approx\frac{2}{\pi}\frac{1}{s\sqrt[3]{n}}\left(\frac{se}{2}\right)^ne^{\pi ns+i(1-n)\frac{\pi}{2}}
\end{equation} 
where
\begin{equation}
s:=\frac{\sum_{p=1}^{n}{s_p}}{n}=\frac{a_H}{4\pi Q_H}\frac{Q_H}{n}=\frac{a_H}{4\pi n}
\end{equation}
such that 
\begin{equation}
I_{\infty}(\{s_p\})\approx\frac{2}{\pi}\frac{1}{s\sqrt[3]{n}}\left(\frac{se}{2}\right)^ne^{\frac{a_H}{4}}e^{i(1-n)\frac{\pi}{2}}
\label{eq:IV.26}
\end{equation} 
Thus, from (\ref{eq:Q2}) we already deduce that in leading order in $a_H$ the entropy is given by the Bekenstein-Hawking law
\begin{equation}
S_{\mathrm{BH}}=\frac{a_H}{4}
\end{equation} 
Since we are counting the number of possible charge-network labels $\lambda_p$, we expect that the lower order to corrections in the entropy do not only depend on the area $a_H$ of the black hole but also on the total charge $Q_H$. We compute these lower order corrections explicitly for the case $G=\mathrm{U}(1)$ in the following section.
\subsubsection{The Abelian case: Maxwell theory}
Let us determine explicitly the lower order corrections to the entropy of the spherically symmetric charged black hole coupled to the Maxwell field. Therefore, we first note that boundary condition (\ref{eq:IV.18}) implies that the signs of the charge quantum numbers $n_p$ and the total charge $Q_H$ need to coincide at every puncture $p\in\mathcal{P}$. Consequently, if we assume w.l.o.g. that $\mathrm{sign}\,Q_H>0$, we have to count for a fixed number $n:=|\mathcal{P}|$ of punctures the number $A(n)$ of nonvanishing charge-network labels $n_p\in\mathbb{N}$, $p\in\mathcal{P}$, such that
\begin{equation}
N=\sum_{p\in\mathcal{P}}{n_p}
\label{eq:IV.28}
\end{equation}
where we wrote $Q_H=Ng$ for some $N\in\mathbb{N}$. Thus, according to (\ref{eq:IV.28}), $A(n)$ just equals the number of partitions of the natural number $N$. This can be computed via elementary combinatorics which yields 
\begin{equation}
A(n)=\begin{pmatrix}
	N-1\\
	n-1
\end{pmatrix}
\label{eq:IV.29}
\end{equation}
Inserting this into (\ref{eq:Q2}) and using (\ref{eq:IV.26}) we find that the total number $\mathcal{N}(a_H,Q_H)$ of admissible surface states is given by  
\begin{equation}
\mathcal{N}(a_H,Q_H)=e^{\frac{a_H}{4}}8\sum_{n=1}^N{\begin{pmatrix}
	N-1\\
	n-1
\end{pmatrix}\left(\frac{e}{8\pi}\right)^nn^{\frac{2}{3}-n}a_H^{n-1}}
\label{eq:IV.30}
\end{equation}
For physical black holes the total area $a_H$ is very large compared to the charge $Q_H$. In this situation, the sum in (\ref{eq:IV.30}) becomes dominated by the highest order term in $a_H$ which is given by $n=N$. We thus find that the entropy of the charged black hole to the next leading order reads
\begin{equation}
S_{\mathrm{BH}}=\frac{a_H}{4}+\left(\frac{Q_H}{g}-1\right)\ln a_H+\ldots
\label{eq:IV.31}
\end{equation}
Hence, we see that the quantum theory yields a logarithmic correction to the Bekenstein-Hawking law with a proportionality factor explicitly depending on the total charge $Q_H$ of the black hole. One can in principle also obtain a result for $P_\text{H}\neq 0$, but the calculation is more complicated and we will not describe it here. 

The result \eqref{eq:IV.31} is quite different from \eqref{eq:3.6.5.9}, \eqref{eq:3.6.5.11} which we have obtained above, in that there is no divergence in the leading order term. For sure, the setup is also different, in that we are working with spherically symmetric horizons here, whereas the direct counting was done for more general geometries. Still, how the difference should be understood remains an open question. We remark that our result \eqref{eq:IV.31} shows some similarity to subleading contributions in the entropy at one loop order in path integral calculations. For example Sen \cite{Sen:2012dw} finds an $\ln A$ correction. His result  for the case of a electrically charged non-rotating black hole in 4-dimensional Einstein gravity reads
\begin{equation}
S_{\mathrm{BH}}=\frac{a_H}{4}+\left(C_1+C_2 \frac{\beta Q^4}{r_H^5}\right)\ln a_H
\end{equation}
where $C_1$ and $C_2$ are numerical constants and the coefficient $\beta$ can be expressed in terms of $a_H$ and $Q_H$ as
\begin{equation}
\beta=2\sqrt{\pi} \frac{a_H^{\frac{3}{2}}}{a_H-2\pi Q_H^2}.
\end{equation}
For small $Q_H$ on finds
\begin{equation}
S_{\mathrm{BH}}\approx \frac{a_H}{4}+\left(C_1+(4\pi)^3
C_2  \frac{Q_H^4}{a_H^2}\right)\ln a_H. 
\end{equation}
The result of Sen is technically an entropy density in terms of mass, whereas the LQG result would arguably be density with respect to area. Transforming our result would add $1/2 \ln a_H$, but that does not change the fact that both results are rather different. Both show that the entropy receives a subleading correction of the form $f(Q_H)\ln a_H$, but the functions of $Q_H$ are quite different. 
\section{Summary and outlook}
\label{se_summ}
In the present work we have studied the classical and quantum theory of isolated horizons of type III that carry YM charges. The main results are:
\begin{itemize}
\item A consistent picture of type III isolated horizons charged under a YM field can be obtained in the quantum theory. 
\item The entropy of the quantum geometry of such a charged isolated horizon is proportional to its area.
\item The degrees of freedom of the YM field can be included in the black hole ensemble, in at least two different ways. In both cases, entropy stays proportional to area. In one of them, a cutoff is needed. 
\item The spherically symmetric limit of a type III charged quantum IH is problematic. 
\end{itemize}
Let us discuss the content in more detail in the following. 
The classical theory is a generalization of \cite{Corichi:2000dm} to type III including the boundary conditions \eqref{eq:3.4.15}, \eqref{eq:3.4.16}. Equation \eqref{eq:3.6.1.15} is also a generalization of the boundary conditions of the gravitational degrees of freedom in the presence of matter fields. 

In the quantum theory, we define the electric charge operator $\widehat{Q}(S)$  \eqref{eq:YM_Q} for Yang-Mills theory with semisimple compact structure group. For the electromagnetic case we sketch a setting in which the horizon can also carry nonzero magnetic charge, and obtain the corresponding magnetic charge operator $\widehat{P}_H$ \eqref{eq:YM_P}. 

We then turn to the quantum theory of type III horizons, and show that the boundary conditions can be implemented in the quantum theory in a straightforward way following the path of \cite{Perez:2010pq}. The operator measuring the distortion at the horizon acquires new terms from the horizon charges, and there are new operators corresponding to the local charge densities \eqref{eq:3.6.3.16}, \eqref{eq:3.6.3.16b}. In the spherical limit, a problem is encountered, since the charge densities are classically fixed to values that can generically not be obtained in the quantum theory. 
This problem can be circumvented for G=U(1) by using an alternative quantization for the electromagnetic field, as described in \eqref{eq_RBohr} ff., but since this solution is particular to the Abelian case, it seems to us that it is more natural to consider the spherically symmetric case as somewhat pathological in the quantum theory.  

The entropy calculation proceeds in the same way as in the uncharged case and leads to the same result, including the same value of the Barbero-Immirzi parameter. However, since we have the YM degrees of freedom at our disposal, we are at liberty to define ensembles that also include YM states at the horizon. For U(1), fluxes can be positive or negative, so we have to introduce a cutoff to keep the result finite. We obtain
\begin{equation}
S_{\mathrm{BH}}(a_H)=\frac{\beta_M^{k,N_{\max}}}{\beta}\frac{a_H}{4l_p^2}-\ln(2N_{\max}), \qquad \beta_M^{k,N_{\max}}=\frac{\ln 2}{\pi\sqrt{3}}+\frac{\ln N_{\max}}{\pi\sqrt{3}}. 
\end{equation}
This result is very similar to what has been found for entanglement entropy across boundaries in matter ground states \cite{Bombelli:1986rw,Srednicki:1993im}, which turns out to be proportional to area, but with a divergent, and hence regulator dependent prefactor. A more recent approach to the
entropy calculation involves analytic continuation \cite{Achour:2014eqa}. In that case, for technical reasons, we are limited to the spherically symmetric case, but this does not cause inconsistencies because the area spectrum becomes continuous and the boundary conditions can be satisfied. 
We find
\begin{equation}
S_{\mathrm{BH}}=\frac{a_H}{4}+\left(\frac{Q_H}{g}-1\right)\ln a_H. 
\end{equation}
The meaning of these results, in particular the question if one of these approaches is the right one, or if both describe different systems, has to be clarified in the future. 

The inclusion of the YM fields also allows for interesting investigations beyond entropy. It had been suggested that there is a correspondence between the quasinormal modes of a Schwarzschild black hole and the area spectrum in LQG \cite{Dreyer:2002vy}. In the present context, it can be asked if there is a similar correspondence between the quasinormal mode frequencies of a Reissner-Nordstrom black hole and the area spectrum. This has been investigated by one of us. However, it appears that there is no such simple connection, except for in the uncharged and in the the extremal limit \cite{KE_master}. 

\acknowledgments
We thank Norbert Bodendorfer and Robert Helling for helpful discussions about some of the material in this work. KE thanks the Elitenetzwerk Bayern and the Elite Masters Course \emph{Theoretical and Mathematical Physics} for financial support. 

\begin{appendix}

\section{Spinor calculus}
\label{ap_spinor}
Let us summarize some important formulas concerning spinor calculus on manifolds which will frequently be used in the main text especially in section \ref{suse_bc}. Let $\{e_I\}$  be an oriented orthonormal tetrad on the Lorentzian manifold $(M,g)$. To this terad, one can associate a corresponding complex null tetrad $(l,k,m,\bar{m})$ as follows
\begin{equation}
l=\frac{e_0-e_3}{\sqrt{2}}\quad k=\frac{e_0+e_3}{\sqrt{2}}\quad m=\frac{e_1+ie_2}{\sqrt{2}}\quad \bar{m}=\frac{e_1-ie_2}{\sqrt{2}}
\label{eq:B5}
\end{equation}
These satisfy the relations
\begin{equation}
l^2=k^2=m^2=\bar{m}^2=l\cdot m=l\cdot\bar{m}=k\cdot m=k\cdot\bar{m}=0\quad\text{and}\quad -l\cdot k=m\cdot\bar{m}=1
\label{eq:B6}
\end{equation}
Assuming that the manifold admits a spin structure, we fix on the corresponding spinor bundle a spinor dyad $(o^A,i^A)$ satisfying $o_Ai^A=1$, where indices are raised and lowered using the antisymmetric symbols $\epsilon_{AB}$ and $\epsilon_{A'B'}$ via $\psi_A:=\psi^B\epsilon_{BA}$ and $\phi^A:=\epsilon^{AB}\phi_B$ for any spinor fields $\psi^A$ and $\phi_A$, respectively. It follows that the real spinor fields of the form $T^{AA'}$ are isomorphic to the ordinary sections $T^{\mu}$ of the tangent bundle. Using the complex null tetrad, such an isomorphism is given by the following soldering form \cite{Ashtekar:1999yj}
\begin{equation}
\sigma^{\mu}_{AA'}:=-i(l^{\mu}i_A\bar{i}_{A'}+k^{\mu}o_A\bar{o}_{A'}-m^{\mu}i_A\bar{o}_{A'}-\bar{m}^{\mu}o_A\bar{i}_{A'})
\label{eq:B7}
\end{equation}
By (\ref{eq:B6}), it satisfies
\begin{equation}
\sigma^{\mu}_{AA'}\sigma_{\nu}^{AA'}=\tensor{g}{^{\mu}_{\nu}},\quad\sigma^{\mu}_{AA'}\sigma_{\mu}^{BB'}=\tensor{\delta}{_A^B}\tensor{\delta}{_{A'}^{B'}}
\label{eq:B8}
\end{equation}
It follows that this soldering form in fact provides such an isomorphism according to $T^{\mu}=\sigma^{\mu}_{AA'}\tensor{T}{^{AA'}}$ and $\tensor{T}{^{AA'}}=\sigma_{\mu}^{AA'}T^{\mu}$, respectively. Applying (\ref{eq:B7}) on the null tetrad, this gives
\begin{equation}
l^{AA'}=io^A\bar{o}^{A'},\,k^{AA'}=ii^A\bar{i}^{A'},\,m^{AA'}=io^A\bar{i}^{A'}\,\bar{m}^{AA'}=ii^A\bar{o}^{A'}
\label{eq:B9}
\end{equation}
Let $\Sigma^{IJ}:=e^I\wedge e^J$ such that we have for the corresponding spinor field  
\begin{equation}
\tensor*{\Sigma}{_{\mu\nu}^{AA'BB'}}=2\tensor{\sigma}{_{[\mu}^{AA'}}\tensor{\sigma}{_{\nu]}^{BB'}}
\label{eq:2.2.1}
\end{equation}
The self-dual part of $\tensor*{\Sigma}{_{\mu\nu}^{AA'BB'}}$ is then given by 
\begin{equation}
\tensor*{\Sigma}{_{\mu\nu}^{AB}}:=\frac{1}{2}\tensor{\bar{\epsilon}}{_{A'B'}}\tensor*{\Sigma}{_{\mu\nu}^{AA'BB'}}=\tensor{\bar{\epsilon}}{_{A'B'}}\tensor{\sigma}{_{[\mu}^{AA'}}\tensor{\sigma}{_{\nu]}^{BB'}}
\label{eq:2.2.2}
\end{equation}
and likewise for the anti self-dual part. Then, since $\tensor{\epsilon}{_{AB}}=o_Ai_B-i_Ao_B$, one can show that $\Sigma^{AB}$ takes the form
\begin{align}
\tensor*{\Sigma}{^{AB}}=k\wedge\bar{m}\,o^Ao^B-\left(m\wedge\bar{m}-l\wedge k\right)o^{(A}i^{B)}-l\wedge m\,i^Ai^B
\label{eq:2.2.3}
\end{align}

\end{appendix}

\end{document}